\documentclass[twocolumn,showkeys,showpacs,article,asp]{revtex4-1}
\usepackage[latin1]{inputenc}
\usepackage{amsmath}
\usepackage{amsfonts}
\usepackage{amssymb}
\usepackage{graphicx}
\usepackage{textcomp}
\usepackage{subfigure}
\usepackage{geometry}
\usepackage{color}

\newcommand*\blue{\color{blue}}

\usepackage[pdfpagemode={UseOutlines},bookmarks=true,bookmarksopen=true,
   bookmarksopenlevel=0,bookmarksnumbered=true,hypertexnames=false,
   colorlinks,linkcolor={blue},citecolor={blue},urlcolor={red},
   pdfstartview={FitV},unicode,breaklinks=true]{hyperref}
\usepackage{multirow}
\usepackage[table,xcdraw]{xcolor}
\begin{document}
\title{Effect of van-Hove singularities in single-walled carbon nanotube leads on transport through double quantum dot system}
\author{Haroon}
\email{haroonjamia@gmail.com}
\author{M.A.H. Ahsan}
\email{mahsan@jmi.ac.in}
\affiliation{Department of Physics, Jamia Millia Islamia (Central University), New Delhi 110025, India}
\date{\today}
\begin{abstract}
The double quantum dot system with single-walled metallic armchair carbon nanotube leads has been studied using Non-equilibrium Green function in the Keldysh formalism. The effect of relative spacing between the energy levels of the dots, interdot tunneling matrix-element, interdot Coulomb interaction and van-Hove singularities in density of states characteristics of quasi-one-dimensional carbon nanotube leads on the conductance of the double quantum dot system has been studied. The conductance and dot occupancies are calculated at finite temperature. It is observed that the density of states of the carbon nanotube leads play a significant role in determining the conductance profile. In particular, whenever the chemical potential of the isolated double quantum dot system is aligned with the position of a van-Hove singularity in the density of states of armchair carbon nanotube leads, the height of the corresponding conductance peak falls considerably. It is further observed that the suppression in the heights of the alternate peaks depends on the relative positions of the energy levels of the dots and their magnitude of separation.
\end{abstract}
\maketitle
\section{Introduction}
\label{sysintro}
\indent
Quantum dots, often termed as artificial atoms, serve as versatile structures that can be used to probe the quantum behavior of electrons on the nanometer scale. In recent times, transport through quantum dots (QD) has been studied extensively both theoretically and experimentally \cite{dkferry,sdatta,jpbird}
leading  to a better understanding of a multitude of underlying physical phenomena  \cite{ilapw,yhasid}
such as the Kondo effect {\cite{kksyc,metkha,rsmsf}}, Coulomb blockade {\cite{ hgmhd, kastner}}, negative differential conductance {\cite{scrkm,jfoe}}, formation of molecular states {\cite{mlldg,jjpph}} etc. Double quantum dots (DQDs) are more promising compared to their single dot counterparts for applications such as quantum information processing, spintronics and quantum computation \cite{dloss,dlevs,hsds,hgbur} etc. The connecting leads to the DQD system 
can be taken to be ideal leads with constant density of states, spin polarized leads or superconducting leads {\cite{mkkiw,crpro,akmza}}. The single walled carbon nanotubes (SWCNTs) {\cite{ijima}} are quasi one-dimensional structures and their density of states(DOS) are characterised by van-Hove singularities(vHs). It is anticipated that these vHs would play significant role in transport through QD systems when SWCNTs are taken as leads. \\
\indent 
Recently, carbon nanotube(CNT) based QDs have been realized \cite{nmmjb, mjbnmj}. A CNT based QD is formed when electrons are confined to a small region within a CNT by the application of gate voltages to the  electrodes, dragging the valence band of the CNT down in energy, thereby causing electrons to pool in a region in the vicinity of the electrode\cite{amjlg}. Experimentally, this is done by laying a CNT on a ${\bf SiO_{2}}$ surface, sitting on a doped ${\bf Si}$ wafer using carbon monoxide through chemical vapor deposition method {\cite{bzclg}}. The ${\bf Si}$ wafer serves as the gate electrode. The CNT can then be connected to metallic leads in order to connect the CNT based QD to an electrical circuit. It is possible to form isolated QDs by induced \cite{phwct,mjbnm} or intrinsic defects {\cite{mbwld}} along the CNT or by tunnel barriers at the metal-CNT interface {\cite{meplb}}. \\
\indent 
The CNT based multiple QDs can also be realized and controlled with the help of electrostatic gates. A pair of metallic top-gates can be used to produce a localized depletion region in the underlying tubes with ohmic contact electrodes. The localized depletion region defines the QD. With the help of top-gates the tunnel barriers and electrostatic energies within single or multiple QDs can also be tuned.
The QDs fabricated this way exhibit familiar characteristics with significant advance in the device control {\cite{mjbsgn}}. Mason {\it et. al.} have measured CNT QDs with multiple electrostatic gates and used the resulting enhanced control to investigate CNT DQDs. Through these device, the transport measurements has revealed honeycomb charge stability diagrams as a function of two independent gate voltages. The tunability of the device allows weak to strong interdot tunnel-coupling regimes and also the leads can be controlled independently. This ability enables one to measure capacitances, energy-level spacings and interaction energies of the system \cite{nmmjb, mjbnmj}.  Chorley {\it et. al.} have also studied tunable Kondo Physics in a CNT based DQD system realized using a SWCNT on a degenerately doped ${\bf Si/SiO_{2}}$ substrate contacted by gold (Au) electrodes. The device uses a central gate to introduce a tunable tunnel barrier, separating the SWCNT into two QDs, which can be controlled individually by additional side gates {\cite{sjcmr}}. 
\\
\indent 
In the present study, we consider a double quantum dot(DQD) system of which one is CNT based, made by artificially creating two tunnel barriers \cite{meplb,mjbnm,mbwld}, and 
 the other dot is tunnel-coupled to the first dot. We assume that parts of CNT remaining after creating a CNT dot within it act as leads to the dot. The leads in our model are thus SWCNTs taken to be metallic. The SWCNTs has a well defined band structure. This way, we have replaced the conventional metallic leads (with flat band structure) by quasi-one-dimensional SWCNT leads with van-Hove singularities. We investigate, how the vHs in the density of states of the SWCNT affect the conductance of the double-quantum dot system.\\
The code for the study has been validated by reproducing the results of single and double quantum dot models as special cases from our model. For example, the results for DQD model in the reference \cite{kps} are reproduced if CNT leads are replaced by the ideal leads (flat band) and the results of single quantum dot model with CNT leads in the reference \cite{sqdcnt} are reproduced in the absence of dot-2 which further reproduces the results of reference \cite{ymnspa} when CNT leads are replaced by ideal leads.\\
\indent
This paper is organized as follows: In Sec. \ref{model} we describe our model with a DQD connected to CNT leads.  Section \ref{eomAndCoupledEqn} gives details of our theoretical formulation based on non-equilibrium  Green functions(NEGFs) and the derivation of DOS of armchair CNTs. In Section \ref{results} we present our  numerical results and Sec. \ref{coclude} the conclusion. 
\section{Model Hamiltonian}
\label{model}
\begin{figure*}
\centering
\caption{Schematic diagram of DQD system in T-shaped geometry. Only dot-1 is coupled to the source and drain SWCNT-leads through  hybridization parameters $V^{s,d}_{k}$. Dots are tunnel-coupled through the matrix-element $t$. The dot energies are given by $\varepsilon_{1}$ and $\varepsilon_{2}$ whereas $U_{1}$ and $U_{2}$ are the respective ondot Coulomb interactions. The parameter $g$ denotes the interdot Coulomb interaction. \hfill\break}
\includegraphics[width=0.85\textwidth]{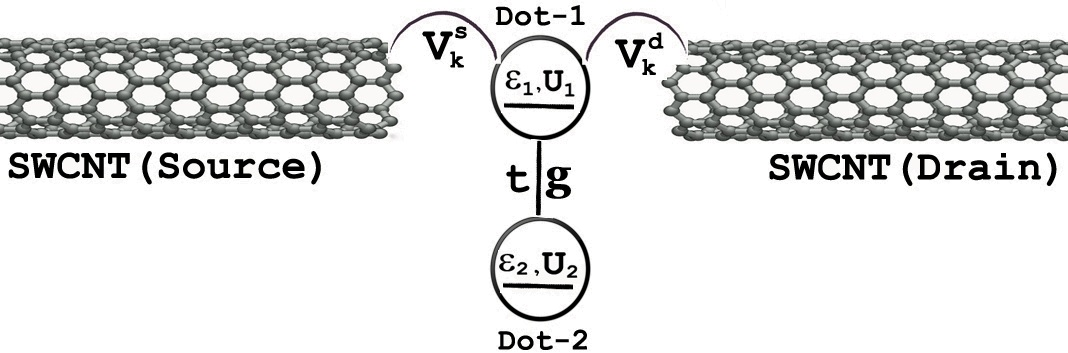} 
\label{fig1model:tshpCNT}
\end{figure*}

The DQDs with SWCNT leads in T-shaped geometry is shown in Fig.\ref{fig1model:tshpCNT}. In T-shaped coupled system, only the central QD (labelled as the dot-1 can be a CNT based QD) is connected to the leads and the other QD (labelled as the dot-2) is connected to the central QD only. 
The system can be described by the  two impurity Anderson type Hamiltonian {\cite{anderson}} consisting of three parts 
\begin{eqnarray}
{\bf H }={\bf H}_{dqd} +{\bf H}_{leads} +{\bf H}_{hyb},
\label{hamilTshpCNT}
\end{eqnarray}
where isolated DQD system is described by the Hamiltonian
\begin{eqnarray*}
{\bf H}_{dqd} &=& \sum_{j=1,2}\varepsilon_{j}\sum_{\sigma} c^{\dagger}_{j\sigma}c_{j\sigma}
+\sum_{j=1,2}U_{j}n_{j\uparrow}n_{j\downarrow}
\nonumber \\ 
&+&g\sum_{\sigma\sigma\prime}n_{1\sigma}n_{2\sigma\prime}
+t\sum_{\sigma}\left(c^{\dagger}_{1\sigma}c_{2\sigma}+ h. c.\right).
\nonumber \\
\label{isolateddqds}
\end{eqnarray*}
The first term in ${\bf H}_{dqd}$ describe energy of an electron on spin degenerate level $\varepsilon_{j}$ of $j^{th}$ dot. Second and third terms containing $U$ and $g$ are many-body terms due to ondot and interdot Coulomb interactions respectively and last term describes interdot tunneling matrix element $t$.
The Hamiltonian ${\bf H}_{leads}$ in eq. (\ref{hamilTshpCNT}) describes the SWCNT-leads with $\varepsilon^{l}_{{\bf k}\sigma}\ (l=s,d)$ as the dispersion relation of a given chirality $(n,m)$ 
\begin{eqnarray*}
{\bf H}_{leads} &=& \sum_{l=s,d}\sum_{{\bf k},\sigma}\varepsilon^{l}_{{\bf k}\sigma}c^{\dagger}_{{\bf k}^{l}\sigma}c_{{\bf k}^{l}\sigma}
\end{eqnarray*}
\begin{eqnarray*}
{\bf H}_{hyb} &=&
\sum_{l=s,d}\sum_{{\bf k},\sigma}\left(V_{\bf k}^{l}c^{\dag}_{{\bf k^{l}}\sigma}c_{1\sigma}+h.c.\right).
\end{eqnarray*}
The Hamiltonian ${\bf H}_{hyb}$ in eq. (\ref{hamilTshpCNT}) describes the coupling of dot-1 with the SWCNT leads. The Hamiltonian ${\bf H}$ in eq. (\ref{hamilTshpCNT}) is $s\leftrightarrow d$ symmetric.
\section{Method}
\subsection{Calculation of Green functions for quantum dots}
\label{eomAndCoupledEqn}
The Green functions (GFs) for QDs are calculated using equation of motion (EOM) method {\cite{lacroix, bulka}}. The EOM method is capable of properly treating the on-site correlations when correlations in the leads are neglected {\cite{ymnspa}}. This method lead to coupled equations incorporating correlation processes to an order restricted by the decoupling scheme. If the number of coupled equations can be handled at even higher order, the higher order correlation processes will be taken into account, leading to enriched physics of the system. In the decoupling procedure we have omitted the higher order correlation processes which lead to electron transfer between the leads and the DQD system. Thus, we have neglected the processes which may occur due to spin flipping or spin quenching processes leading to the effects such as the AbrikosovSuhl resonance or the Kondo effect {\cite{bulka, pscd}}. The scheme, however, properly incorporates charge fluctuations and all correlations in the DQD system.\\
\indent 
The retarded GFs for the dots in EOM method are obtained as follows
\begin{eqnarray}
G_{j}^{r}\left(t^{\prime},t\right) &=& \left<\left< c_{j}\left( t\right),c_{j}^{\dag}\left( t^{\prime}\right) \right>\right>^{r}
\nonumber \\
&=&-i\,\Theta\left(t-t^{\prime}\right)  \left< \left\lbrace c_{j}\left( t\right),c_{j}^{\dag}\left( t^{\prime}\right) \right\rbrace\right> 
\label{retardgreen}
\end{eqnarray}
where $\Theta(t-t^{\prime})$ is the step function, for convenience one can fix $t^{\prime}=0$. The GF in the above eq. ({\ref{retardgreen}) is differentiated with respect to time $t$, which generates higher order GFs. The EOM for these new GFs is to be set up without any approximation to any term or GFs. In the next iteration of EOM one finds further higher order GFs, this process continues unless one implements some decoupling scheme. The Hartree-Fock approximation \cite{zheng} is one such decoupling procedure that do not keep more than two operators in the GFs \cite{zimin}. The Hartree-Fock approximation treats the system in a effective non-interacting way as the many-particle GFs due to many-body interactions are decoupled into one-particle GFs. 
In our calculation, two-particle GFs (containing upto four operators) are retained as they are and the higher ones are decoupled into two-particle GFs as $ \left<\left< n_{j{\bar{\sigma}}}\left(t\right)n_{\bar{j}\sigma}\left(t\right)c_{j\sigma}\left(t\right), c^{\dag}_{j\sigma}\left(0\right) \right> \right>^{r} = \left<  n_{j{\bar{\sigma}}}\right>\left<\left<n_{\bar{j}\sigma}\left(t\right)c_{j\sigma}\left(t\right), c^{\dag}_{j\sigma}\left(0\right)\right>\right>^{r} + \left< n_{\bar{j}\sigma}\right>\left<\left<n_{j\bar{\sigma}}\left(t\right)c_{j\sigma}\left(t\right), c^{\dag}_{j\sigma}\left(0\right)\right>\right>^{r}$ where $j\left(\bar{j}\right)=1\left(2\right)$. The GFs such as $\left<\left< c^{\dag}_{k^{s(d)}\sigma(\bar{\sigma})}\left(t\right)c_{k^{s(d)}\bar{\sigma}(\sigma)}\left(t\right)c_{j\sigma(\bar{\sigma})}\left(t\right), c^{\dag}_{j\sigma(\bar{\sigma})}\left(0\right)\right>\right>^{r}$ vanish as the correlations within leads and between the dots and the leads are neglected {\it i.e.} $\left< c^{\dag}_{k^{s(d)}\sigma} c_{k^{s(d)}\sigma} \right> = 0 $ and  $\left< c^{\dag}_{j\sigma}c_{k^{s(d)}\bar{\sigma}} \right> = 0 $.
In this way a set of eight coupled equations for each dots are obtained. The set of coupled equations for the GFs $G^{r}_{j\sigma}(\omega)$  are not $1\leftrightarrow 2$ symmetric as the Hamiltonian is not $1\leftrightarrow 2$ symmetric. However, the coupled equations, if calculated to even higher order will be $s\leftrightarrow d$ symmetric as the Hamiltonian ${\bf H}$ in eq. (\ref{hamilTshpCNT}) is $s\leftrightarrow d$ symmetric. The set of eight coupled equations for dot-1 are given as
\begin{widetext}
\begin{subequations}
\begin{eqnarray}
A_{1}
G_{11\sigma}^{r}(\omega) = 1+tG_{21\sigma}^{r}(\omega)
+ U_{1}G_{41\sigma}^{r}(\omega)
+g\left(G_{31\sigma}^{r}(\omega)+G_{51\sigma}^{r}(\omega)\right)
\label{1tcntdot1}
\end{eqnarray}
\begin{eqnarray}
A_{2}G_{21\sigma}^{r}(\omega) = tG_{11\sigma}^{r}(\omega)
+
U_{2}G_{61\sigma}^{r}(\omega)
+g\left(G_{71\sigma}^{r}(\omega)+G_{81\sigma}^{r}(\omega)\right)
\label{1tcntdot2}
\end{eqnarray}
\begin{eqnarray}
A_{3}G_{31\sigma}^{r}(\omega) =  \left<n_{2\sigma}\right> 
+U_{1}\left<n_{2\sigma}\right>G_{41\sigma}^{r}(\omega)
+
g\left<n_{2\sigma}\right>G_{51\sigma}^{r}(\omega)
+tG_{71\sigma}^{r}(\omega)
\label{1tcntdot3}
\end{eqnarray}
\begin{eqnarray}
A_{4}G_{51\sigma}^{r}(\omega) = \left<n_{2\bar{\sigma}}\right> + tG_{61\sigma}^{r}(\omega)
+U_{1}\left<n_{2\bar{\sigma}}\right>G_{41\sigma}^{r}(\omega)
+
g\left<n_{2\bar{\sigma}}\right>G_{31\sigma}^{r}(\omega)
\label{1tcntdot5}
\end{eqnarray}
\begin{eqnarray}
A_{5}G_{61\sigma}^{r}(\omega) = 
tG_{51\sigma}^{r}(\omega)
+g\left<n_{2\bar{\sigma}}\right> \left(G_{71\sigma}^{r}(\omega)+G_{81\sigma}^{r}(\omega)\right)
\label{1tcntdot6}
\end{eqnarray}
\begin{eqnarray}
A_{6}G_{41\sigma}^{r}(\omega) =\left<n_{1\bar{\sigma}}\right>
+tG_{81\sigma}^{r}(\omega)
+g\left<n_{1\bar{\sigma}}\right> \left(G_{31\sigma}^{r}(\omega)+G_{51\sigma}^{r}(\omega)\right)
\label{1tcntdot4}
\end{eqnarray}
\begin{eqnarray}
A_{7}G_{71\sigma}^{r}(\omega) = U_{2}\left<n_{1\sigma}\right>G_{61\sigma}^{r}(\omega)
+g\left<n_{1\sigma}\right>G_{81\sigma}^{r}(\omega)
+ tG_{31\sigma}^{r}(\omega)
\label{1tcntdot7}
\end{eqnarray}
\begin{eqnarray}
A_{8}G_{81\sigma}^{r}(\omega) = tG_{41\sigma}^{r}(\omega)
+U_{2}\left<n_{1\bar{\sigma}}\right>G_{61\sigma}^{r}(\omega)
+g\left<n_{1\bar{\sigma}}\right>G_{71\sigma}^{r}(\omega).
\label{1tcntdot8}
\end{eqnarray}
\label{eqnsetfordot1}
\end{subequations}
\end{widetext}
Similar eight equations for dot-2 can also be found. In the above equations 
$G_{(1/2)1\sigma}^{r}(\omega)=\left<\left<c_{(1/2)\sigma}\left(\omega\right);c^{\dag}_{1\sigma}\left(\omega^\prime\right)\right>\right>^{r}_{\omega}$,
$G_{(3/4)1\sigma}^{r}(\omega)= \left<\left<n_{(2/1)(\sigma/\bar{\sigma})}\left(\omega\right)c_{1\sigma}\left(\omega\right);c^{\dag}_{1\sigma}\left(\omega^\prime\right)\right>\right>^{r}_{\omega}$,
$G_{(5/6)1\sigma}^{r}(\omega)= \left<\left<n_{2\bar{\sigma}}\left(\omega\right)c_{(1/2)\sigma}\left(\omega\right);c^{\dag}_{1\sigma}\left(\omega^\prime\right)\right>\right>^{r}_{\omega}$,
 $G_{(7/8)1\sigma}^{r}(\omega)= \left<\left<n_{1(\sigma/\bar{\sigma})}\left(\omega\right)c_{2\sigma}\left(\omega\right);c^{\dag}_{1\sigma}\left(\omega^\prime\right)\right>\right>^{r}_{\omega}$
with $A_{1}= \left(\omega-\varepsilon_{1}-\Sigma_{\sigma}^{r}\left(\omega \right)\right)$, 
$A_{2}= \left(\omega-\varepsilon_{2}\right)$,
$A_{3(4)}= \left(A_{1}-g- U_{1}\left<n_{1\bar{\sigma}}\right>-g\left<n_{2\bar{\sigma}(\sigma)}\right> \right)$, 
 $A_{5(6)}= \left(A_{2(1)}- U_{2(1)}-g\left<n_{1(2)\sigma}\right>-g\left<n_{1(2)\bar{\sigma}}\right>\right)$ and 
$A_{7(8)}= \left(A_{2}- g-U_{2}\left<n_{2\bar{\sigma}}\right>-g\left<n_{1\bar{\sigma}(\sigma)}\right>\right)$.
The self-energies arising due to coupling of dot-1 to the SWCNT-leads are evaluated as 
\begin{eqnarray}
\Sigma_{\sigma}^{r}\left(\omega \right)=\sum_{k} \left| V_{k}^{s(d)} \right|^{2}\frac{1}{\omega -\varepsilon_{k\sigma}^{s(d)}}
\label{selfenergeiscnt}
\end{eqnarray}
with $\omega\rightarrow\omega^{+}=\omega+i\delta$, we have
\begin{eqnarray}
\Sigma_{\sigma}^{s(d)}&&\left(\omega^{+} \right)
\nonumber \\
=&& \Re\left[\Sigma_{\sigma}^{s(d)}\left(\omega^{+} \right) \right]-i\pi\left| V^{s(d)} \right|^{2}\rho_{CNT}(\omega).
\label{selfenergeiscnt}
\end{eqnarray}
The imaginary part of the self-energy $\Im\left[\Sigma_{\sigma}^{s(d)}\left(\omega^{+} \right) \right]$ simplifies to $-\Gamma^{s(d)}\rho_{CNT}(\omega)$, where $\rho_{CNT}(\omega)$ is the density of states of SWCNT leads. In calculation we have neglected the real-part of the self energy $\Re\left[\Sigma_{\sigma}^{s(d)}\left(\omega^{+} \right) \right]$ as it only causes the dot-level shifting {\cite{pekka}}.
\subsection{Calculation of the density of states of armchair carbon nanotubes}
\label{dosarm}
The DOS of SWCNTs can be calculated from their energy dispersion relation which can be obtained from the dispersion relation of the graphene {\cite{dressel}}. The aspect ratio (ratio of the CNT length along the tube axis to the diameter) of a SWCNT is high. The structure of a SWCNT is macroscopic along the tube axis with infinite number of states along it but, the number of states in the circumferential direction will be quantized thus limited to a finite number. The nearest neighbour tight-binding energy dispersion relation for graphene is given as {\cite{reich,srctjm}}
\begin{eqnarray}
E\left(k_{x},k_{y}\right)=\pm t_{0}\left[1+4\alpha\cos\left(\frac{\sqrt{3}k_{x}a}{2}\right)+4\alpha^{2}\right]^{\frac{1}{2}} \nonumber \\
\label{dispgraphene}
\end{eqnarray}
where $\alpha=\cos\left(\frac{k_{y}a}{2}\right)$ and $a=0.246 nm$ is the in-plane lattice constant for the graphene and $t_{0}$ is nearest-neighbour hopping element. Following the exact analytical derivation for the energy-dispersion relation and the density of states for SWCNTs in references \cite{akinwande,dejiakin}, we briefly present here, the necessary steps for obtaining DOS for armchair SWCNTs and the positions of the characteristic vHs in the DOS.\\ 
\indent
 Rolling up graphene sheet along the chiral vector ${\bf C}_{h}=n{{\bf a}_{1}}+m{\bf a}_{2}$ where $n$ and $m$ are the chirality indices ($m=n$ for armchair CNT) and ${\bf a}_{1}=\left(\frac{\sqrt{3}}{2},\frac{1}{2} \right)a$, ${\bf a}_{2}=\left(\frac{\sqrt{3}}{2},-\frac{1}{2} \right)a$ are the real space unit vectors of the hexagonal lattice {\cite{dressel}},  we have periodic boundary condition ${\bf C}_{h}\cdot {\bf K} = 2\pi q$ leading to
\begin{eqnarray}
k_{x}=\frac{2\pi q}{\sqrt{3}n a}
\label{kx}
\end{eqnarray} 
where $q=1\cdots 2n$ is the subband index. Substituting eq. (\ref{kx}) in eq. ({\ref{dispgraphene}), we obtain dispersion relation for the armchair CNT as
\begin{eqnarray}
E_{q}^{AC}\left(k\right)=\pm t_{0}\left[ 1+4\alpha^{\prime}\cos\left(\frac{\pi q}{n}\right)+4{\alpha^{\prime}}^{2} \right]^{\frac{1}{2}}
\label{dispArmchair}
\end{eqnarray}
where $\alpha^{\prime}=\cos(\frac{ka}{2})$ and $-\pi<ka<\pi$. In an ideal one-dimensional structures, the DOS is $\frac{1}{\pi}\left|\frac{dK}{dE}\right|$ where E is the energy and $K$ the wave vector {\cite{dejiakin}}. The DOS per subband for the armchair CNT is thus given as
\begin{eqnarray}
\rho(E,q) =\frac{4\tau}{a\pi}\frac{|E|}{E^{\prime}\sqrt{(E^{\prime}-A_{c1})(-E^{\prime}+A_{c2})}}.
\label{dossubnadAC}
\end{eqnarray}
Where $\tau=1(2)$ at the center(otherwise) is the zone degeneracy of the Brillouin zones. Assuming CNT to be infinitely long, the DOS (normalized per unit length) for the armchair CNT $(n,n)$ can be written as 
\begin{eqnarray}
\rho_{AC}(E)&=&\sum_{q=1}^{2n}\rho(E,q).
\label{dosAC}
\end{eqnarray}
The DOS depends upon the chirality index $n$. The parameters in eq.(\ref{dossubnadAC}) are given as $E^{\prime}=\sqrt{E^{2}-V_{hAC}^{2}}$, $V_{hAC}=\pm\left|t_{0}\sin{\left(\frac{\pi q}{n}\right)} \right|$, $A_{c1}=t_{0}\left(-2+\cos\left(\frac{\pi q}{n}\right) \right)$
and $A_{c2}=t_{0}\left(2+\cos\left(\frac{\pi q}{n}\right) \right)$. 
The vHs define the energy space where the DOS of CNTs is finite and real. They are present due to the quasi-sinusoidal energy dispersion relation in eq. (\ref{dispArmchair}). Their positions can be obtained from the denominator of $\rho_{AC}$ in eq. (\ref{dosAC}).\\
\indent
Consider $\left(\sqrt{E^{2}-V_{hAC}^{2}}-A_{c1} \right)=0$  
substituting for $V_{hAC}$ and $A_{c1}$ the positions of vHs are given as
\begin{eqnarray}
E=\pm t_{0}\sqrt{5-4\cos\left(\frac{\pi q}{n} \right)}.
\label{vanhoveAC2}
\end{eqnarray}
The other factors $\left( E^{2}-V_{hAC}^{2}\right)^{1/2}$ and $\left(-\sqrt{E^{2}-V_{hAC}^{2}}+A_{c2}\right)$ lead to positions of vHs as 
\begin{eqnarray}
E=\pm t_{0}\sin\left(\frac{\pi q}{n} \right)
\label{vanhoveAC3}
\end{eqnarray}
and 
\begin{eqnarray}
E=\pm t_{0}\sqrt{5+4\cos\left(\frac{\pi q}{n} \right)}
\label{vanhoveAC4}
\end{eqnarray}
respectively.
The energies given by the eqs. (\ref{vanhoveAC2}), (\ref{vanhoveAC3}) and (\ref{vanhoveAC4}) are the positions of the vHs. However, eqs. (\ref{vanhoveAC2}) and (\ref{vanhoveAC4}) give same vHs for $q=1\cdots 2n$, therefore any of these equations can be used with eq. (\ref{vanhoveAC3}) to find exact number of vHs positions. 
\begin{table}
\caption{Positions of vHs in the DOS of the armchair $(5,5)$ CNT obtained using eqs. (\ref{vanhoveAC2}) and (\ref{vanhoveAC3}).}
\begin{ruledtabular}
\begin{tabular}{
>{\columncolor[HTML]{FFFFFF}}c |
>{\columncolor[HTML]{FFFFFF}}c |
>{\columncolor[HTML]{FFFFFF}}c |
>{\columncolor[HTML]{FFFFFF}}c 
}
\multicolumn{4}{c}{\cellcolor[HTML]{FFFFFF}}       
 \\
\multicolumn{4}{c}{\multirow{-2}{*}{\cellcolor[HTML]{FFFFFF}{Armchair $(5,5)$ SWCNT}}} \\ \hline
{S. No} & {Singularity} & {S. No.} & {Singularity}    \\ \hline
{1}     & {-2.8699}     & {8}      & {\color[HTML]{3531FF} {0.5887}}   \\ \hline
{2}     & {-2.4972}     & {9}      & {0.9511}                         \\ \hline
{3}     & {-1.9401}     & {10}     & {1.0000}                          \\ \hline
{4}     & {-1.3281}     & {11}     & {1.3281}                         \\ \hline
{5}     & {-1.0000}     & {12}     & {1.9401}                          \\ \hline
{6}     & {-0.9511}     & {13}     & {2.4972}                           \\ \hline
{7}     & {-0.5887}     & {14}     & {2.8699}                        
\end{tabular}
\end{ruledtabular}
\label{tab:sigularities}
\end{table}
\begin{figure}
\caption{Density of states (DOS) per unit cell vs. Energy/$t_{0}$ for the armchair $(5,5)$ CNT. The positive and negative energies correspond to the conduction and valance bands, respectively. The vertical sharp lines represent vHs. The first vHs lies at E=0.5887, there are total 14 vHs in the band symmetric about the Fermi energy at $E=E_{f}=0.0$. The finite values of DOS about the Fermi energy signifies metallic nature of the armchair $(5,5)$ CNT. \hfill \break}
\centering
\includegraphics[width=0.45\textwidth]{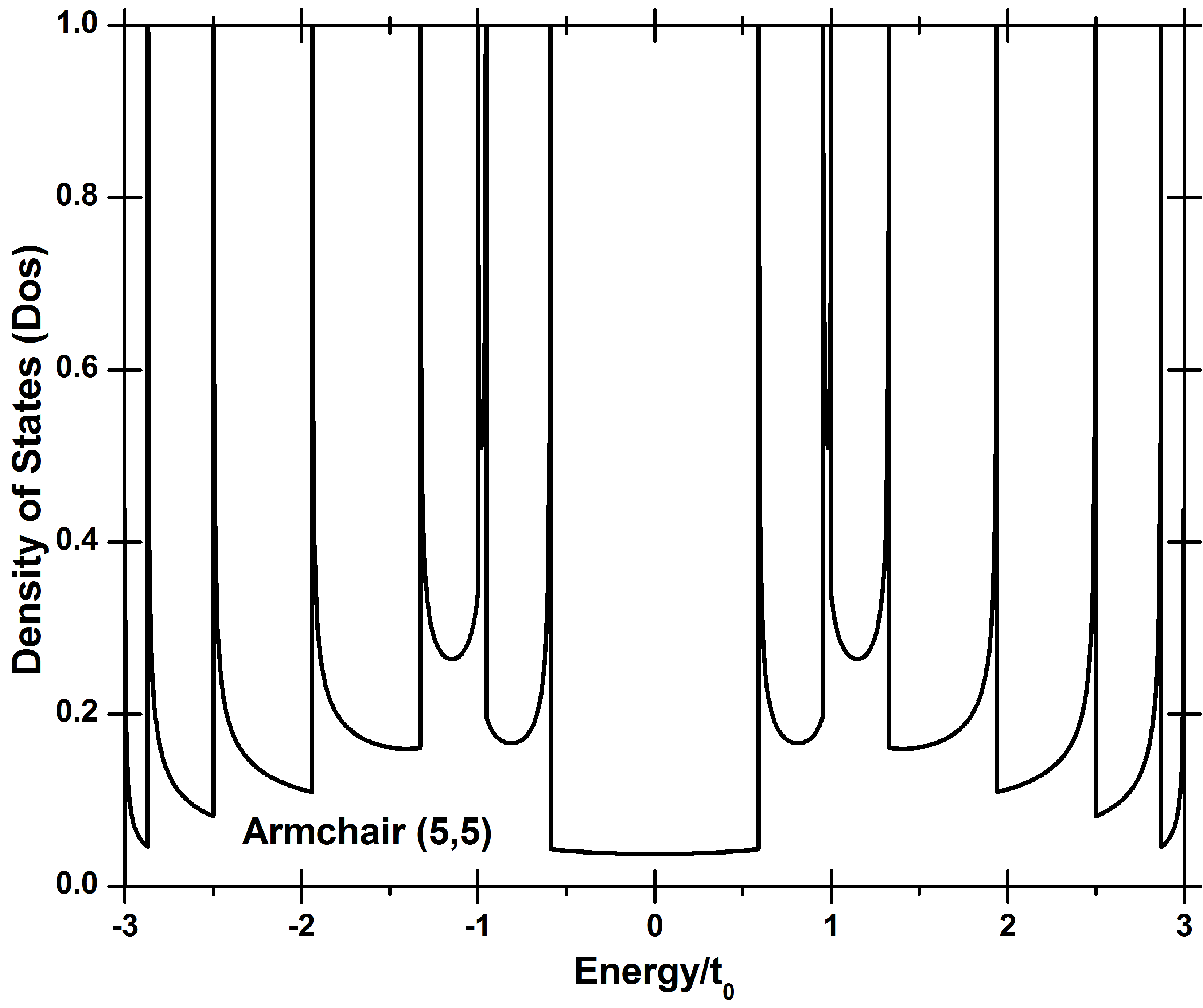} 
\label{dos55}
\end{figure}
\\
\indent
Using the explicit expression for the DOS of armchair $(n,n)$ SWCNT obtained in eq. (\ref{dosAC}), we have plotted in Fig. \ref{dos55} the DOS for armchair $(5,5)$ CNT. The sharp vertical lines represent the positions of vHs. The region between the positions of first vHs about the Fermi energy in the valance and conduction band has nearly flat structure. All armchair CNTs are metallic since the DOS at the Fermi energy is finite. There are 14 vHs present in the DOS of armchair $(5,5)$ CNT given in Table {\ref{tab:sigularities}}. 
\subsection{Calculation of conductance and dot occupancies}
\label{condAndOccup}
An expression for the current given in eq. (\ref{currentforfinitetemp}) can be obtained using Keldysh NEGF formalism \cite{ymnsw} whence the conductance at finite temperature can be obtained
\begin{eqnarray}
J=&&-\frac{e}{\hbar}\int_{-\infty}^{\infty}d\omega\, \left( f_{s}(\omega)-f_{d}(\omega) \right) 
\nonumber \\
\times && tr\,\left[\frac{{\bf \Gamma}^{s}(\omega){\bf \Gamma}^{d}(\omega)}{{\bf \Gamma}^{s}(\omega)+{\bf \Gamma}^{d}(\omega)}\left\lbrace -\frac{1}{\pi}\Im[{\bf G}^{r}(\omega)] \right\rbrace \right].
\label{currentforfinitetemp}
\end{eqnarray} 
In above eq. (\ref{currentforfinitetemp}), $f_{s(d)}\left(\omega\right)=\frac{1}{1+e^{\beta\left(\omega-\mu_{s(d)}\right)}}$ is the Fermi distribution function in the source (drain) lead. 
Where $\mu_{s}=\epsilon_{F}+eV$ and $\mu_{d}=\epsilon_{F}$ are the chemical potentials of the source and drain leads respectively, such that $\mu_{s}-\mu_{d}=+eV$, $e$ is the electronic charge and $\beta^{-1}=k_{B}T$. 
The current in eq. (\ref{currentforfinitetemp}) is differentiated with respect to the biasing voltage $V$. The zero bias conductance can be obtained as ${\cal G}=\left.\frac{dJ}{dV}\right|_{V\rightarrow 0}$. In this case  chemical potentials of the source and drain leads are varied simultaneously. Tracing out the product of coupling matrices ${\bf \Gamma^{s(d)}}$ and retarded GFs ${\bf G}^{r}(\omega)$, the final form of the conductance formula for the DQD system is given as
\begin{eqnarray}
{\cal G}=&&\frac{e^{2}}{h}\sum_{\sigma}\int_{-\infty}^{+\infty}d\omega f_{eq}^{'}\left(\omega\right)
\nonumber \\
\times && \frac{2\Gamma_{1 \sigma}^{s}\left(\omega\right)\Gamma_{1 \sigma}^{d}\left(\omega\right)}{\Gamma_{1 \sigma}^{s}\left(\omega\right)+\Gamma_{1\sigma}^{d}\left(\omega\right)}Im\left[G^{r}_{1\sigma}\left(\omega\right) \right]
\label{zerobiasConduc}
\end{eqnarray}
where the couplings between the dot-1 and SWCNT leads are given by 
\begin{eqnarray}
\Gamma_{1\sigma}^{s(d)}\left(\omega\right)=2\pi\sum_{k}\left|V_{ k\sigma}^{s(d)}\right|^{2}\delta\left( \omega-\epsilon_{k\sigma}^{s(d)}\right). 
\label{couplingdotcnt}
\end{eqnarray} 
In numerical calculation the hybridization parameters $V_{ k\sigma}^{s(d)}$ can taken to be spin $\sigma$ and energy $k$ independent {\cite{dasilva,ding,lopez}}. The finite-temperature dot occupancies are calculated as the integral of local interacting DOS $\rho_{j}\left(\omega\right) = -\frac{1}{\pi}\Im\left[ G^{r}_{j\sigma}\left(\omega\right)\right]$ weighted by the Fermi function $f_{eq}\left(\omega\right)$ {\cite{ymnspa,kps}} given as
\begin{eqnarray}
\left\langle n_{j\sigma}\right\rangle = -\frac{1}{\pi}\int_{-\infty}^{+\infty}d\omega f_{eq}\left(\omega\right)\Im\left[G^{r}_{j\sigma}\left(\omega\right)\right].
\label{finitetempoccup}
\end{eqnarray}
The $f_{eq}\left(\omega\right)=\frac{1}{1+e^{\beta\left(\omega-\mu\right)}}$ is the equilibrium Fermi-distribution function and $f^{'}_{eq}\left(\omega\right)$ is its derivative. The conductance in eq. (\ref{zerobiasConduc}) also depends GF of the dot-2 through occupancies $\left<n_{i\sigma} \right>$, $i=1,2$   in eq. (\ref{eqnsetfordot1}) which are to be calculated in a self consistent manner. 
\section{Numerical results and discussion}
\label{results}
In the following, we present a systematic study with armchair $(5,5)$ CNT to investigate the effect of vHs present in its DOS on the conductance of the system by varying chemical potential for different relative energy level spacings of the dots $|\varepsilon_{1}-\varepsilon_{2}|$ and interdot tunneling matrix-element $t$. The conductance is calculated in the Coulomb blockade regime {\it i.e.} the ondot Coulomb interaction 
$U_{1}=U_{2}=U$ is the largest energy scale in the system. All energies are measured in units of $U$. The coupling to the leads are very small ($\Gamma^{s(d)} \ll 1$, weak coupling regime).
The temperature $k_{B}T$ is the smallest energy parameter so that the smearing of the conductance peaks caused by the temperature does not mask the effects of other system parameters. We consider here, the paramagnetic case i.e. $<n_{j\sigma}>=<n_{j\bar{\sigma}}>$. 
In Fig. {\ref{condcnt1}}, we plot the conductance as a function of chemical potential $\mu$ with  armchair $(5,5)$ CNT leads for two different values of interdot tunneling matrix-element $t=0.07U,0.1U$. The heights and respective positions of the peaks are shown as (Position, Height). The dot levels are taken as $\left|\varepsilon_{1}-\varepsilon_{2}\right|\,> k_{B}T$ and the interdot Coulomb interaction 
as $g=0.08U$ {\it i.e.} small compared to the ondot Coulomb interaction $U$. In the weak coupling regime $\Gamma\ll 1$, the many-body eigenstates of the isolated DQD system are not much affected. Therefore, a conductance peak is seen every time as the chemical potential $\mu$ in the leads aligns with the difference $\lambda_{N+1}^{0}-\lambda_{N}^{0}$, where $\lambda_{N}^{0}$ is the $N-$electron ground state of the isolated DQDs. This is called the resonant tunneling {\cite{tknpa}}.
\begin{figure}[h]
\centering
\includegraphics[width=0.45\textwidth]{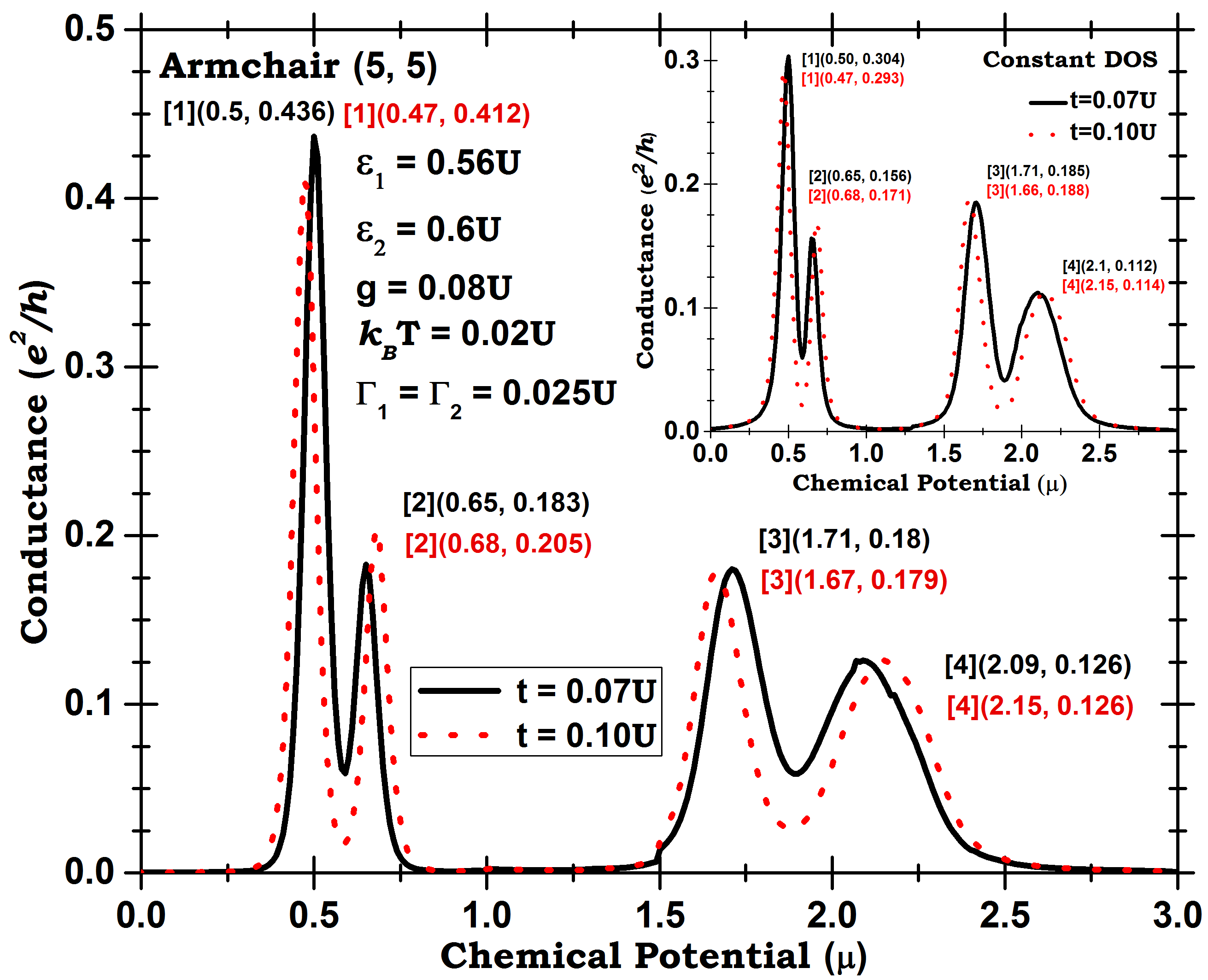}
\caption{{Conductance (in units of $e^{2}/h$) {\it versus}  chemical potential $\mu$ (in units of $U$) for armchair $(5,5)$ CNT. The plots correspond to two different values of interdot tunneling matrix-element: $t=0.07U-$solid line; $t=0.1U-$dashed line. Other parameters have been taken as $\varepsilon_{1}=0.56U$, $\varepsilon_{2}=0.6U$, $g=0.08U$, $k_{B}T=0.02U$ and $\Gamma_{1}=\Gamma_{2}=0.025U$. In the inset, the conductance of the DQD system with armchair $(5,5)$ CNTs replaced by the ideal leads (constant DOS) is shown for the same values of parameters.\hfill \break}}
\label{condcnt1}
\end{figure}
In Fig. {\ref{condcnt1}}, it is observed that when the chemical potential $\mu$ is far below the one-electron ground state energy of the isolated DQD system, the conductance is zero. As the chemical potential $\mu$ approaches close to value of the one-electron ground state energy, resonant tunneling takes place and a peak in the conductance is seen. When the chemical potential increases beyond this value the conductance is again zero due to Coulomb blockade {\cite{tknpa}}. 
Successive peaks in the conductance are seen whenever chemical potential overcomes the Coulomb blockade situation.
 If dot-2 is not present in the model, there would be only two peaks appearing at $\epsilon$ and $\epsilon+U$. In presence of dot-2, each peak split into two peaks due to interdot tunneling matrix-element $t$. When $U>>g$ and $g\approx 0$, the splitting between the first and second peak is roughly given by $\sqrt{(\varepsilon_{1}-\varepsilon_{2})^{2}+4t^{2}}$. 
 For two values of interdot tunneling matrix-element $t=0.07U,0.1U$, it is observed from Fig {\ref{condcnt1}} that as the interdot tunneling matrix-element $t$ increases, the splitting between the first and second peak increases. The splitting between the  third and fourth peaks is also affected by increasing $t$ values. The interdot tunneling matrix-element $t$ significantly affects the heights of the first two peaks. The broadening of the peaks gradually increases from the first to fourth peak due to ondot and interdot Coulomb interactions. All these observations are similar to those with ideal leads in earlier studies {\cite {pals,lamba,kps}}, but it is observed that the heights of the peaks are higher with CNT leads compared to ideal leads. For comparison we have also plotted inset Fig. \ref{condcnt1} for same values of parameters with leads having constant DOS. This is due to the metallic nature of the DOS of armchair $(5,5)$ CNT. 
\begin{figure}[!htb]
\subfigure[\, $\varepsilon_{1}=0.56U$, $\varepsilon_{2}=0.589U$]{\includegraphics[width=0.9\linewidth]{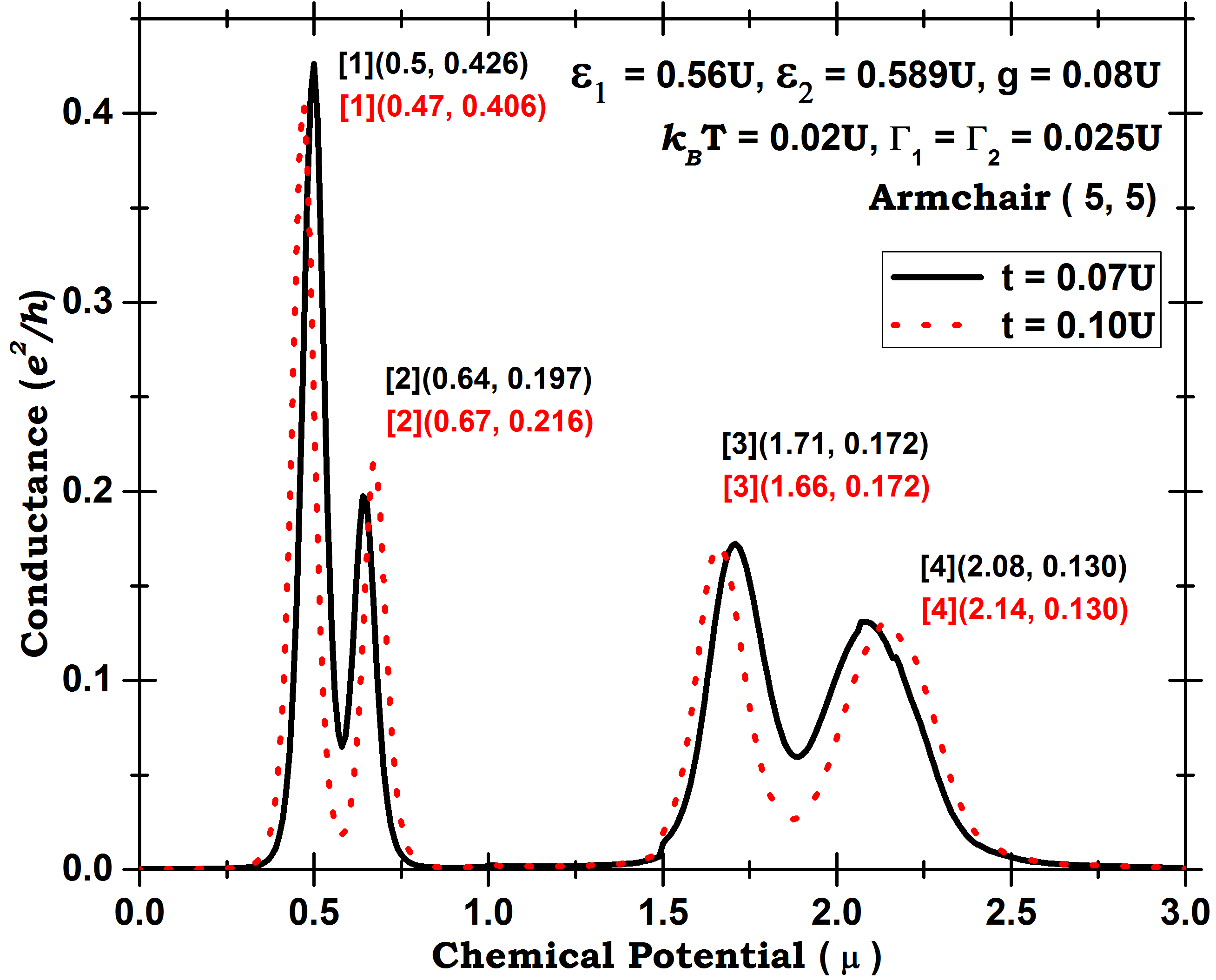}\label{singatdot2}}
\subfigure[\, $\varepsilon_{1}=0.589U$, $\varepsilon_{2}=0.56U$]{\includegraphics[width=0.9\linewidth]{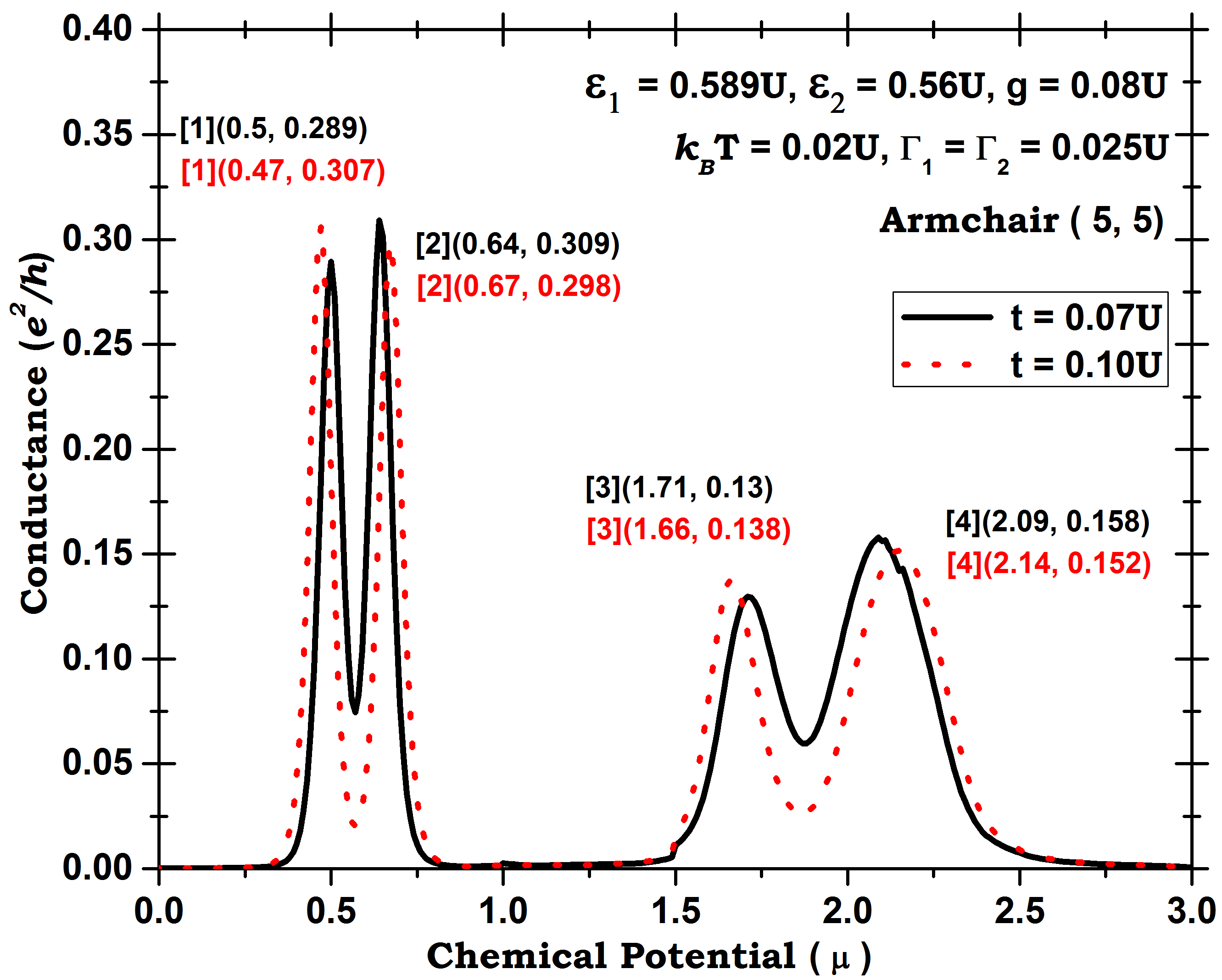}\label{singatdot1}} 
\caption{Conductance (in units of $e^{2}/h$) {\it versus} chemical potential $\mu$ (in units of $U$) for armchair $(5,5)$ CNT. The plots correspond to two different values of interdot tunneling matrix-element: $t=0.07U-$solid line; $t=0.1U-$dotted line in (a) for $\varepsilon_{1}=0.56U$, $\varepsilon_{2}=0.589U$. (b) for $\varepsilon_{1}=0.589U$, $\varepsilon_{2}=0.56U$. Other parameters are taken as $g=0.08U$, $k_{B}T=0.02U$ and $\Gamma_{1}=\Gamma_{2}=0.025U$.\hfill \break}
\label{onedotAtsigularity1}
\end{figure}
\\
\indent  
We now study, the effect of swapping energy levels of the dots. In Fig. {\ref{onedotAtsigularity1}}, we plot conductance as a function of chemical potential $\mu$ in the armchair $(5,5)$ CNT leads for two different values of the interdot tunneling matrix-element $t=0.07U,0.1U$. The energy levels of the dots, in Figs. {\ref{singatdot2}} and {\ref{singatdot1}} are respectively kept as $\varepsilon_{1}=0.56U$, $\varepsilon_{2}=0.589U$ and $\varepsilon_{1}=0.589U$, $\varepsilon_{2}=0.56U$ with interdot Coulomb interaction $g=0.08U$. In either case, $\left|\varepsilon_{1}-\varepsilon_{2}\right|=0.029U$ and the energy level of one of the two dots is aligned with the first vHs about the Fermi level in the DOS of the armchair $(5,5)$ CNT leads at $0.589$. The ground states of the isolated DQD system with $U_1=U_2=U$ corresponding to one, three and four electrons are respectively given as $\lambda^{0}_{1e}=\frac{1}{2}\left[\varepsilon_1+\varepsilon_2-\sqrt{(\varepsilon_1-\varepsilon_2)^2+4t^2}\right]$,  $\lambda^{0}_{3e}=\frac{1}{2}\left[3\left(\varepsilon_{1}+\varepsilon_{2}\right)+4g+2U-\sqrt{(\varepsilon_{1}-\varepsilon_{2})^2+4t^2}\right]$, and $\lambda^{0}_{4e}=2(\varepsilon_{1}+\varepsilon_{2})+4g+2U$, whereas for two electrons  $\lambda^{0}_{2e}$ is obtained numerically. At zero temperature, the four peaks would appear at the positions $\lambda^{0}_{1e}=0.5030$, $\lambda^{0}_{2e}-\lambda^{0}_{1e}=0.7051$, $\lambda^{0}_{3e}-\lambda^{0}_{2e}=1.6039$ and $\lambda^{0}_{4e}-\lambda^{0}_{3e}=1.8060$, respectively. But the positions of peaks in Fig. {\ref{onedotAtsigularity1}} are slightly shifted due to several factors such as the temperature, ondot and interdot interactions also coupling to the leads. The actual positions of respective four peaks for $t=0.07U$, as read from Figs. {\ref{singatdot2}} or {\ref{singatdot1}} are at $0.50$, $0.64$, $1.70$ and $2.08$. The splitting between any two corresponding peaks in Figs. {\ref{singatdot2}}} and {\ref{singatdot1}} remains the same, but the heights of the peaks are different. It is seen that with $\varepsilon_{1}<\varepsilon_{2}$ in Fig. {\ref{singatdot2}}, there is alternate suppression in the peak heights, {\it i.e.} the heights of the second and fourth peaks are less as compared to the first and third peaks, respectively. Whereas for $\varepsilon_{1}>\varepsilon_{2}$ in Fig. {\ref{singatdot1}}, the heights of the first and third peaks are less as compared to the second and fourth peaks, respectively. In either case, the first and second peaks exhibit significant change in their heights as compared to the third and the fourth peak heights. With this observation, one may attribute the alternate suppression of peak heights to the effect of aligning one of the dot levels with the position of the vHs in the DOS of CNT leads. This we further investigate in the following.  
\begin{figure}
\subfigure[\, $\varepsilon_{1}=0.971U$, $\varepsilon_{2}=1.0U$]{\includegraphics[width=0.9\linewidth]{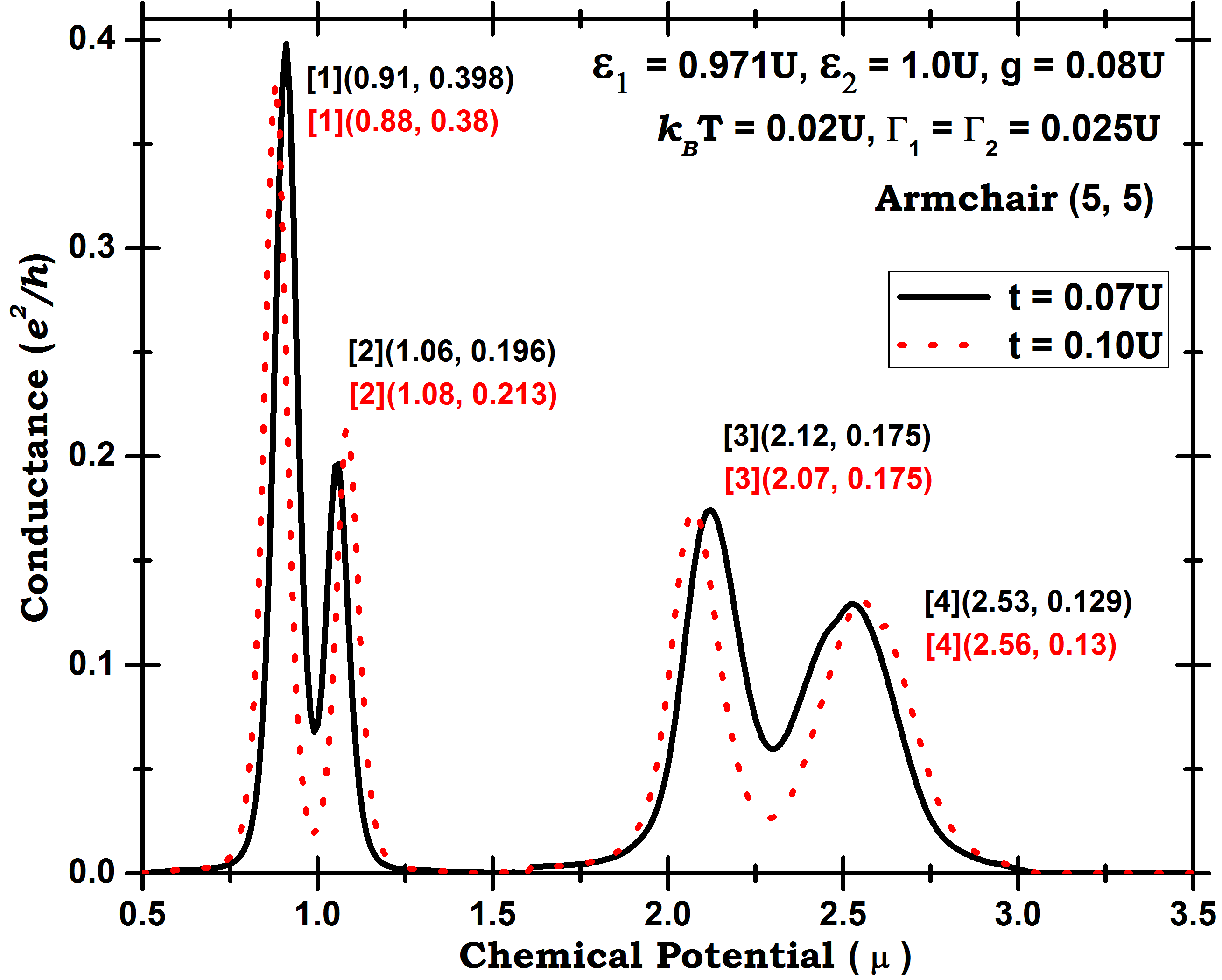}\label{sing2atdot2}}
\subfigure[\, $\varepsilon_{1}=1.0U$, $\varepsilon_{2}=0.971U$]{\includegraphics[width=0.9\linewidth]{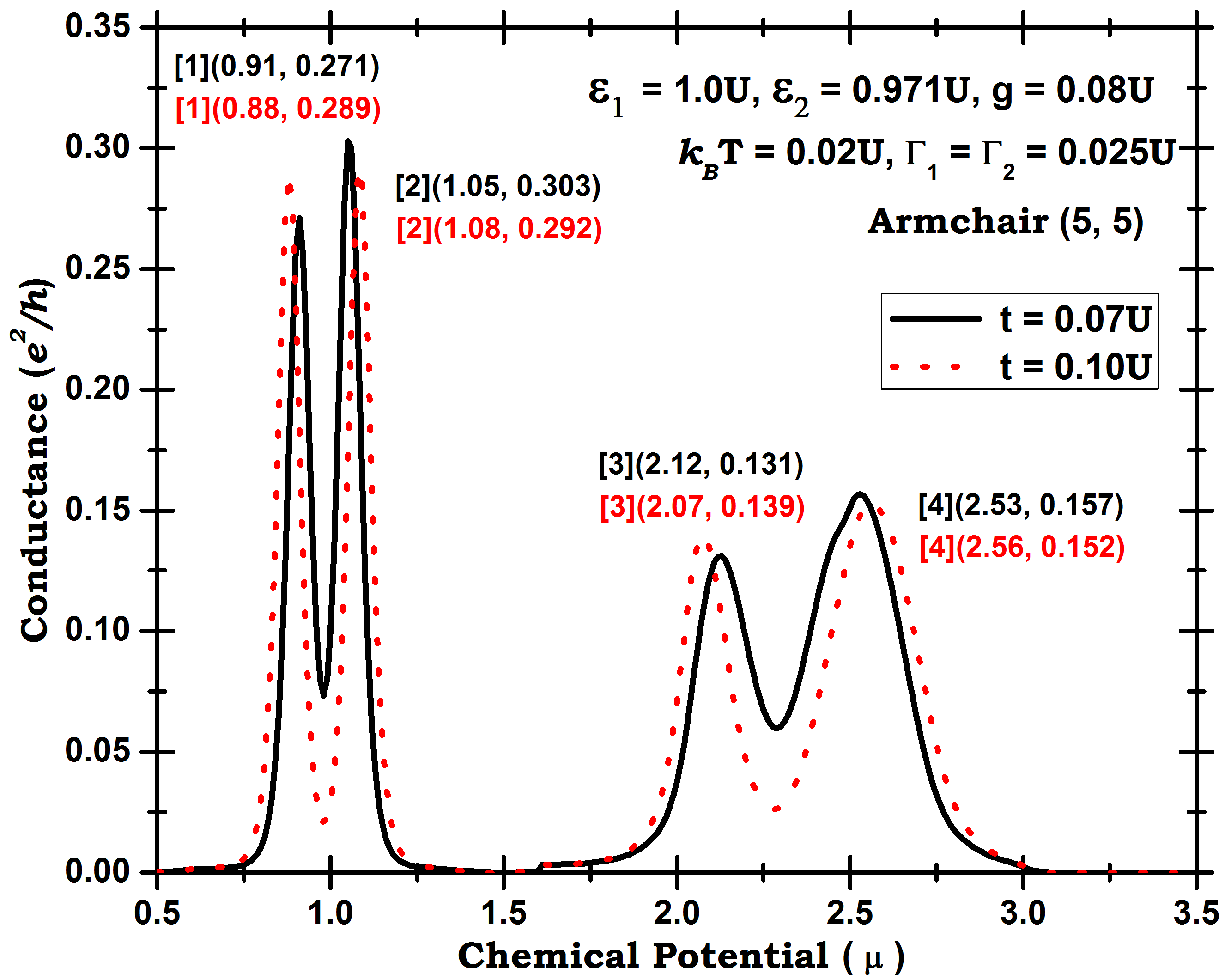}\label{sing2atdot1}} 
\caption{Conductance (in units of $e^{2}/h$) {\it versus}  chemical potential $\mu$ (in units of $U$) for armchair $(5,5)$ CNT. The plots correspond to two different values of interdot tunneling matrix-element: $t=0.07U-$solid line; $t=0.1U-$dashed line in (a) for $\varepsilon_{1}=0.971U$, $\varepsilon_{2}=1.0U$. (b) for $\varepsilon_{1}=1.0U$, $\varepsilon_{2}=0.971U$. Other parameters are taken as $g=0.08U$, $k_{B}T=0.02U$ and $\Gamma_{1}=\Gamma_{2}=0.025U$.}
\label{onedotAt2ndsigularity}
\end{figure}
\\
\indent
To examine it further, in Fig. {\ref{onedotAt2ndsigularity}}, we plot conductance as a function of chemical potential $\mu$ in the armchair $(5,5)$ CNT leads for same values of parameters as in Fig. {\ref{onedotAtsigularity1}}, except that  the dot levels in Fig. {\ref{sing2atdot2}}, are kept at values $\varepsilon_{1}=0.971U$, $\varepsilon_{2}=1.0U$ and in Fig. {\ref{sing2atdot1}}, at values $\varepsilon_{1}=1.0U$, $\varepsilon_{2}=0.971U$. The energy level of one of the two dots, in either case is aligned with the second vHs position; away from the Fermi level in the DOS of the armchair $(5,5)$ CNT at $1.0$. The dot levels are kept separated by the same value $\left|\varepsilon_{1}-\varepsilon_{2}\right|=0.029$ as in Fig. {\ref{onedotAtsigularity1}}. It is observed from Figs. {\ref{onedotAtsigularity1}} and {\ref{onedotAt2ndsigularity}} that the conductance profiles in the two situations resemble, qualitatively. The difference lies only in the heights of peaks which are less in Fig. {\ref{onedotAt2ndsigularity}} compared to the ones in Fig. {\ref{onedotAtsigularity1}}. This is due to the effect of variation in DOS of CNT leads. 
\begin{figure}
\subfigure[\, $\varepsilon_{1}=\varepsilon_{2}=0.589U$]{\includegraphics[width=0.9\linewidth]{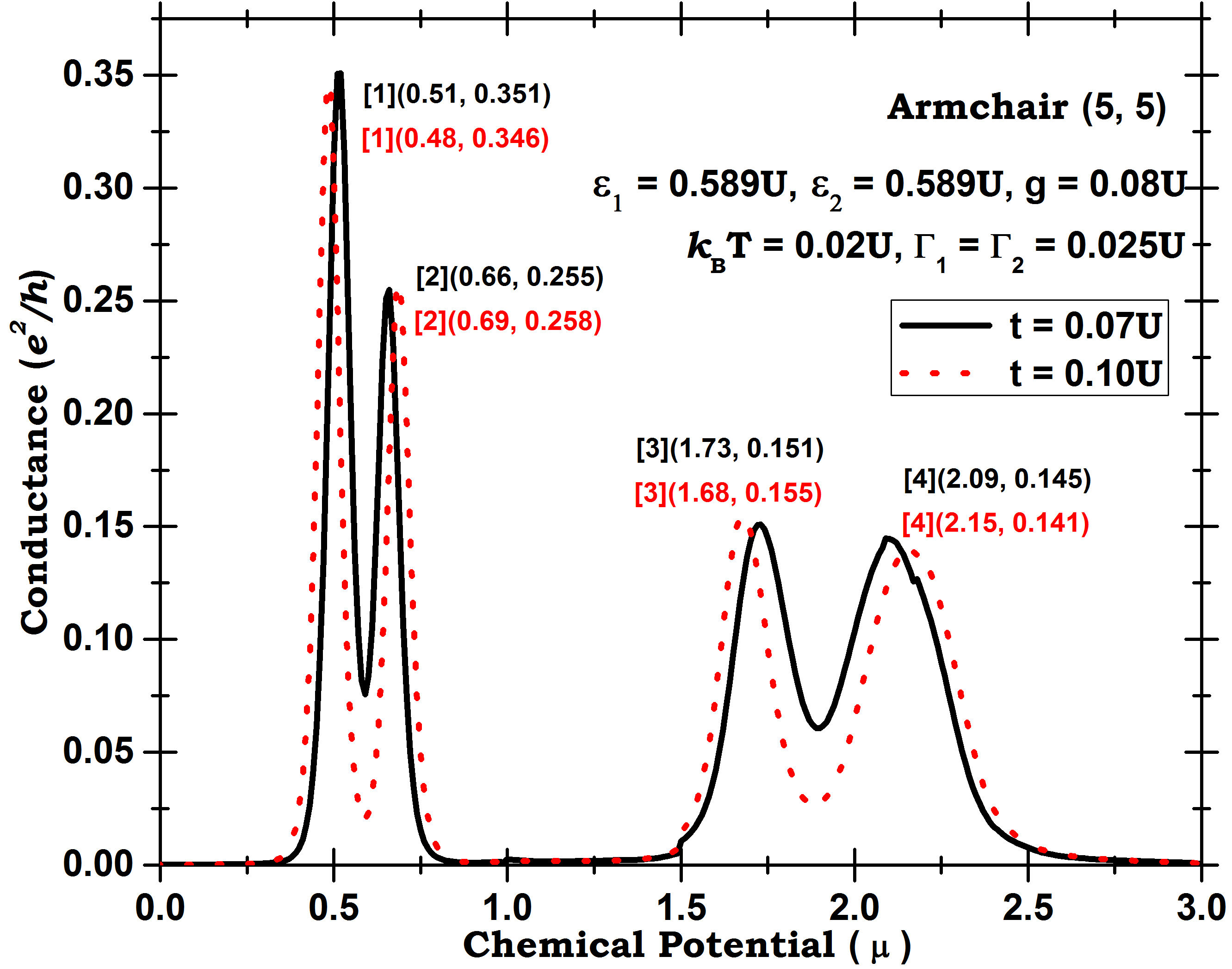}\label{2dotatsing1}}
\subfigure[\, $\varepsilon_{1}=\varepsilon_{2}=0.10U$]{\includegraphics[width=0.9\linewidth]{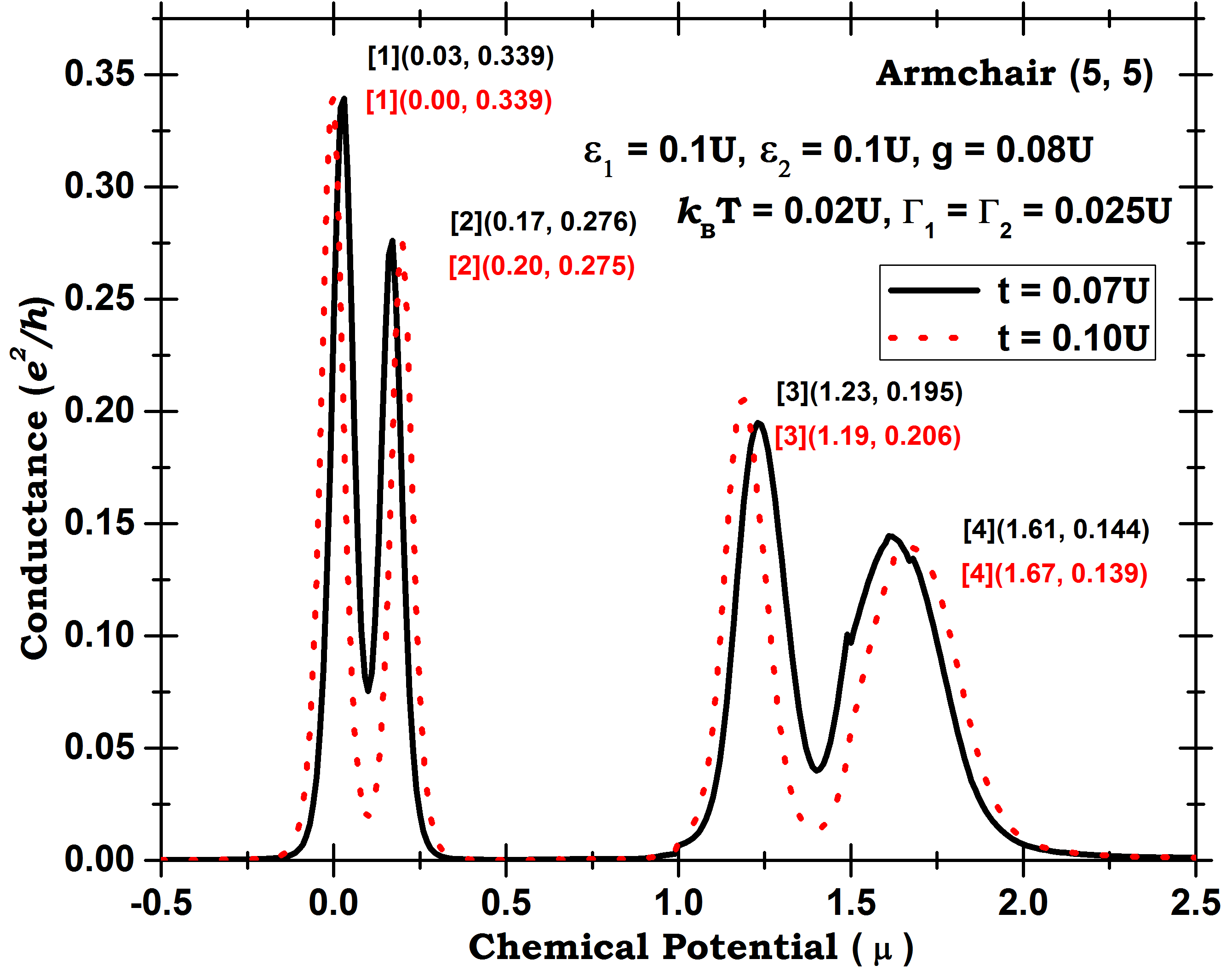}\label{2dotaway4rmsing1}}
\caption{Conductance (in units of $e^{2}/h$) {\it versus}  chemical potential $\mu$ (in units of $U$) for armchair $(5,5)$ CNT. The plots correspond to two different values of interdot tunneling matrix-element: $t=0.07U-$solid line; $t=0.1U-$dotted line in (a) for $\varepsilon_{1}=\varepsilon_{2}=0.589U$. (b) for $\varepsilon_{1}=\varepsilon_{2}=0.1U$. Other parameters are taken as $g=0.08U$, $k_{B}T=0.02U$ and $\Gamma_{1}=\Gamma_{2}=0.025U$.}
\label{twodotat1stSingandAway}
\end{figure}
\\
\indent
In Fig. {\ref{twodotat1stSingandAway}}, we plot conductance as a function of  chemical potential $\mu$ in the armchair $(5,5)$ CNT leads for two different values of the interdot tunneling matrix-element $t=0.07U,0.1U$. The dot levels in Fig. {\ref{2dotatsing1}}, are fixed to align with the first vHs position present near the Fermi level in the DOS of the armchair $(5,5)$ CNT at $0.589$ {\it i.e.} $\varepsilon_{1}=\varepsilon_{2}=0.589U$ and in Fig. {\ref{2dotaway4rmsing1}}, at same values closed to the Fermi level in the leads $\varepsilon_{1}=\varepsilon_{2}=0.10U$. A rough approximation of the positions of peaks, at zero temperature, can be obtained by the ground states of the isolated DQD system. The positions of all four peaks in this case can be obtained with $\varepsilon_{1}=\varepsilon_{2}=\varepsilon$ and $U_{1}=U_{2}=U$ from the results given above. The positions of the first, second, third and fourth peaks are given as $E_{0}^{1}=\varepsilon-t$, $E_{0}^{2}=\varepsilon+t+\frac{1}{2}\left(g+U- \sqrt{(U-g)^{2}+16t^{2}}\right)$, $E_{0}^{3}=(\varepsilon-t)+\frac{1}{2}(3g+U)+\frac{1}{2}\sqrt{(U-g)^{2}+4t^{2}}$ and $E_{0}^{4}=\varepsilon+t+2g+U$, respectively {\cite{bulka}}. Using peak positions; peak separations $E_{0}^{i}-E_{0}^{j}$ between the peaks can also be obtained \cite{haizhou}. The corresponding positions of the peaks obtained using above expressions for $t=0.07$, in Fig. {\ref{2dotatsing1}, are given as $0.5190$, $0.7182$, $1.6198$ and $1.8190$ whereas in Fig. {\ref{2dotaway4rmsing1}}, at $0.03$, $0.2292$, $1.1308$ and $1.33$.  However, the actual positions are slightly different. It is observed that the heights of the peaks gradually fall and their widths broaden from the first to fourth peak. This is due to the electron-electron interaction which modifies each time an electron is added when $N-$electron chemical potential of isolated DQD system reached. The first peak corresponds to non-interacting case and is sharpest and tallest. This picture modifies when there are two electrons in the DQD system. Now, the two electrons can interact in two ways, they can stay on the same dot by costing an additional energy equal to the ondot Coulomb interaction $U$ or on different dot by costing an additional energy equal to the interdot Coulomb interaction $g$. Further, the three electron situation on DQD would necessarily cost an additional energy $U$ and also $g$ as one of the two dots must be doubly occupied. The four electron situation has only possible configuration that each dot is doubly occupied, necessitating an energy $U$ and $g$ to be further incorporated. The gradual peak broadening and falling from first to fourth peak is therefore caused by the electron-electron correlation at a fixed temperature. 
\\
\indent
It is observed that the heights of the second and third peaks in Fig. {\ref{2dotatsing1}} are shorter as compared to the heights of the corresponding peaks in Fig. {\ref{2dotaway4rmsing1}} whereas, the first and fourth peaks are nearly of the same heights. The DOS of the CNT leads slightly differs about the positions of the first and fourth peaks in two situations, but significantly differs about the second and third peaks positions. 
\\
\indent
The conductance profiles when energy levels of both the dots are kept at values $\varepsilon_{1}=\varepsilon_{2}=0.951U$ and $\varepsilon_{1}=\varepsilon_{2}=1.0U$ are found similar to those in Fig. {\ref{2dotatsing1}}. The DOS of CNT has vHs at these values as given in Table {\ref{tab:sigularities}. The conductance profiles when energy levels of both the dots are kept at values $\varepsilon_{1}=\varepsilon_{2}=0.953U$ and $\varepsilon_{1}=\varepsilon_{2}=0.952U$ were observed to be  almost identical as DOS of CNT vary very slightly at the positions of conductance peaks in two cases. These values lie close the vHs at $0.951$ in DOS of CNT; the conductance profiles in two cases are qualitatively similar to the above cases for $\varepsilon_{1}=\varepsilon_{2}$.

\begin{figure}
\subfigure[\, $\varepsilon_{1}=0.59$, $\varepsilon_{2}=0.4U$]{\includegraphics[width=0.9\linewidth]{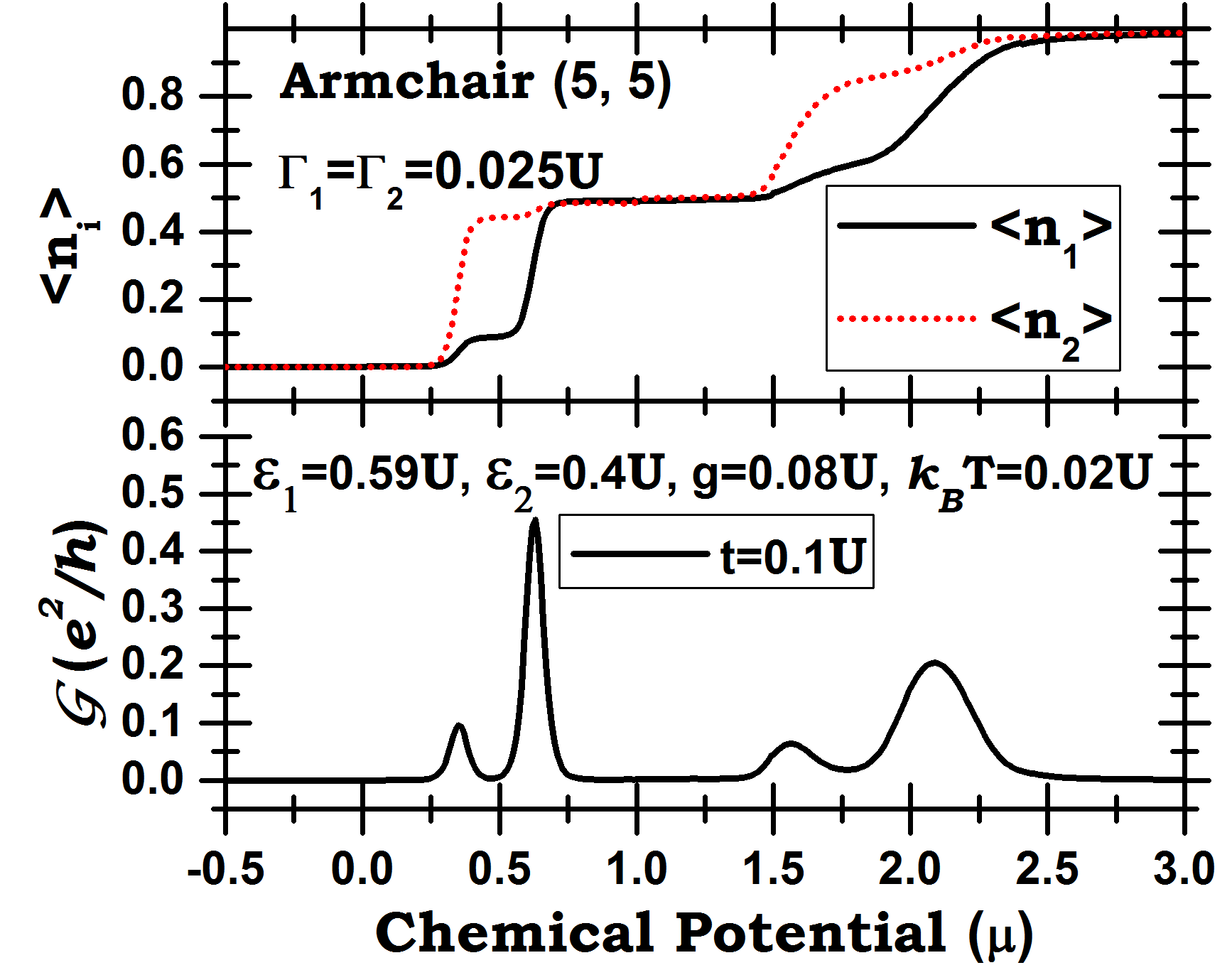} \label{sing1e1ge2}}
\subfigure[\, $\varepsilon_{1}=0.4$, $\varepsilon_{2}=0.59U$]{\includegraphics[width=0.9\linewidth]{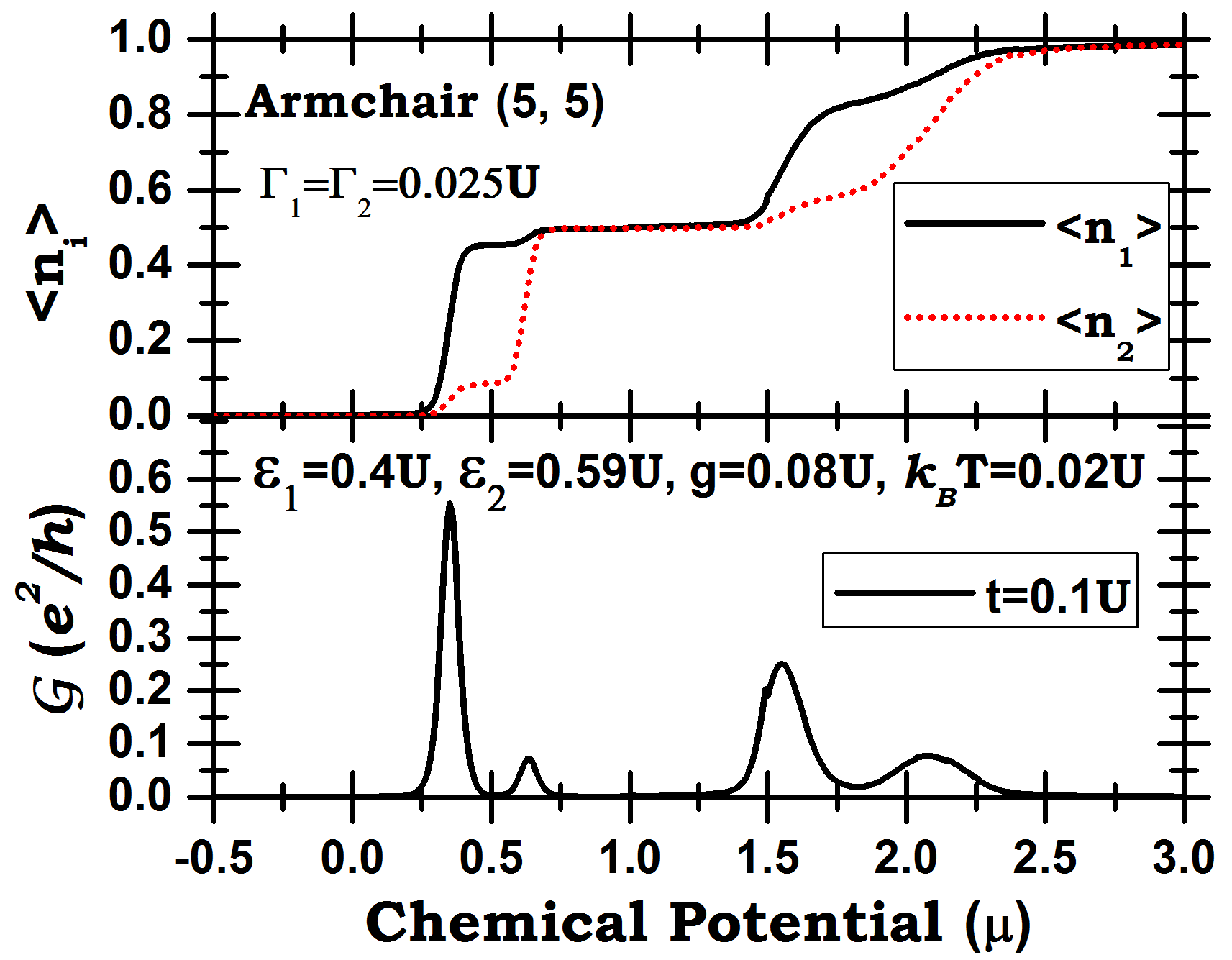} \label{sing1e2ge1}}
\caption{Conductance (in units of $e^{2}/h$) and corresponding occupancies of the dots $<n_{i}>$ {\it versus}  chemical potential $\mu$ (in units of $U$) for armchair $(5,5)$ CNT. The plots correspond at a fixed value of the interdot tunneling matrix-element $t=0.1U$ in (a) for $\varepsilon_{1}=0.59$, $\varepsilon_{2}=0.4U$. (b) for $\varepsilon_{1}=0.4$, $\varepsilon_{2}=0.59U$. Other parameters are taken as $g=0.08U$, $k_{B}T=0.02U$ and $\Gamma_{1}=\Gamma_{2}=0.025U$.}
\label{largeseparation}
\end{figure}
\indent
To investigate the effect of large separation between the energy levels of the dots, in Fig. {\ref{largeseparation}}, we plot conductance and the corresponding occupancies $<n_i>$ of the dots as a function of  chemical potential $\mu$ in the armchair $(5,5)$ CNT leads at a fixed value of the interdot tunneling matrix-element $t=0.1U$ with $g=0.08U$ and $k_{B}T=0.02U$. The dot levels in Fig. {\ref{sing1e1ge2}}, are kept at values $\varepsilon_{1}=0.59U$, $\varepsilon_{2}=0.4U$ and in Fig. {\ref{sing1e2ge1}}, at $\varepsilon_{1}=0.4$, $\varepsilon_{2}=0.59U$. The dot levels are far separated by value $\left|\varepsilon_{1}-\varepsilon_{2}\right|=0.19U$. One of the two dots in either case is aligned with the first vHs position away from the Fermi level; present in the DOS of armchair $(5,5)$ CNT leads at $0.589$. It is observed that the alternate conductance peaks are suppressed similar to those seen in Figs. {\ref{onedotAtsigularity1}} and {\ref{onedotAt2ndsigularity}}, but the differences between the heights of the suppressed and unsuppressed peaks are relatively large. The corresponding occupancies in Figs. {\ref{sing1e1ge2}} and {\ref{sing1e2ge1}}, show plateau structures investigated as a function of  chemical potential. The structure originates due the quantization of charge. The occupancies show that the peaks in the conductance profiles in Figs. {\ref{sing1e1ge2}} and {\ref{sing1e2ge1}}, correspond to the jumps in the occupancies of dot-1 $\left< n_{1}\right>$. The sharp jump signifies narrow peak where the tapered as broad peak. In Fig. {\ref{sing1e1ge2}}, corresponding to the first peak, the occupancy of the dot-1 is less than that of dot-2 {\it i.e.} $\left< n_1\right> < \left< n_2\right>$, due to the fact that the energy level of dot-2 is fixed below that of the dot-1 {\it i.e.} $\varepsilon_{2}<\varepsilon_{1}$. There are only two pathways possible for electron transport in the system viz. $Source\rightarrow Dot-1\rightarrow Drain$ ($S D_{1} D$) and $Source\rightarrow Dot-1\rightarrow Dot-2\rightarrow Dot-1\rightarrow Drain$ ($S D_{1}D_{2}D_{1}D$). However, the conductance results due to the transport through the quantum mechanical superposition of these two paths {\cite{scgrkc,scrkm}}. In one-electron situation with $\varepsilon_{2}<\varepsilon_{1}$, the path $S D_{1}D_{2}D_{1}D$ is more probable. In the two-electron situation, the second electron occupies dot-1 in order to avoid extra energy due to ondot Coulomb interaction $U$ therefore, the corresponding occupancy of dot-1 is more than that dot-2 {\it i.e.} $\left< n_{1}\right> > \left< n_{2}\right>$. In this situation, the path $S D_{1} D$ is more probable and there is a sharper jump in the occupation number of dot-1 $\left< n_{1}\right>$ leading to a narrow peak with long height. The third electron would prefer to occupy the dot-2, since $U$ (not $g$) will make a difference in the energy as $(2\varepsilon_{2}+U)<(2\varepsilon_{1}+U)$. The occupancy of dot-2 is therefore, more than that of dot-1 $\left< n_2\right> > \left< n_1\right>$, the path $S D_{1}D_{2}D_{1}D$ is more probable. There is a tapered jump in the occupation number of dot-1 $\left< n_1\right>$ due to electron-electron correlation leading to broad and short peak. Now, the fourth electron has only one possibility to occupy the first dot and to take the path $S D_{1} D$. The occupancy of dot-1, which is more than that of the dot-2 $\left< n_1\right> > \left< n_2\right>$, has a tapered jump due to correlation, lead to a peak further broader in the width. The peaks present in the conductance profile in Fig. {\ref{sing1e2ge1}}, for $\varepsilon_{1}<\varepsilon_2$ has similar behavior to that in Fig. {\ref{sing1e1ge2}}. In Fig. {\ref{largeseparation}}, one of the two dots in either case is aligned with the vHs position, but suppression in the heights of peaks are not due to vHs position rather due to large dot level separations.  
\begin{figure}
\subfigure[\, $\varepsilon_{1}> \varepsilon_{2}$]{\includegraphics[width=0.9\linewidth]{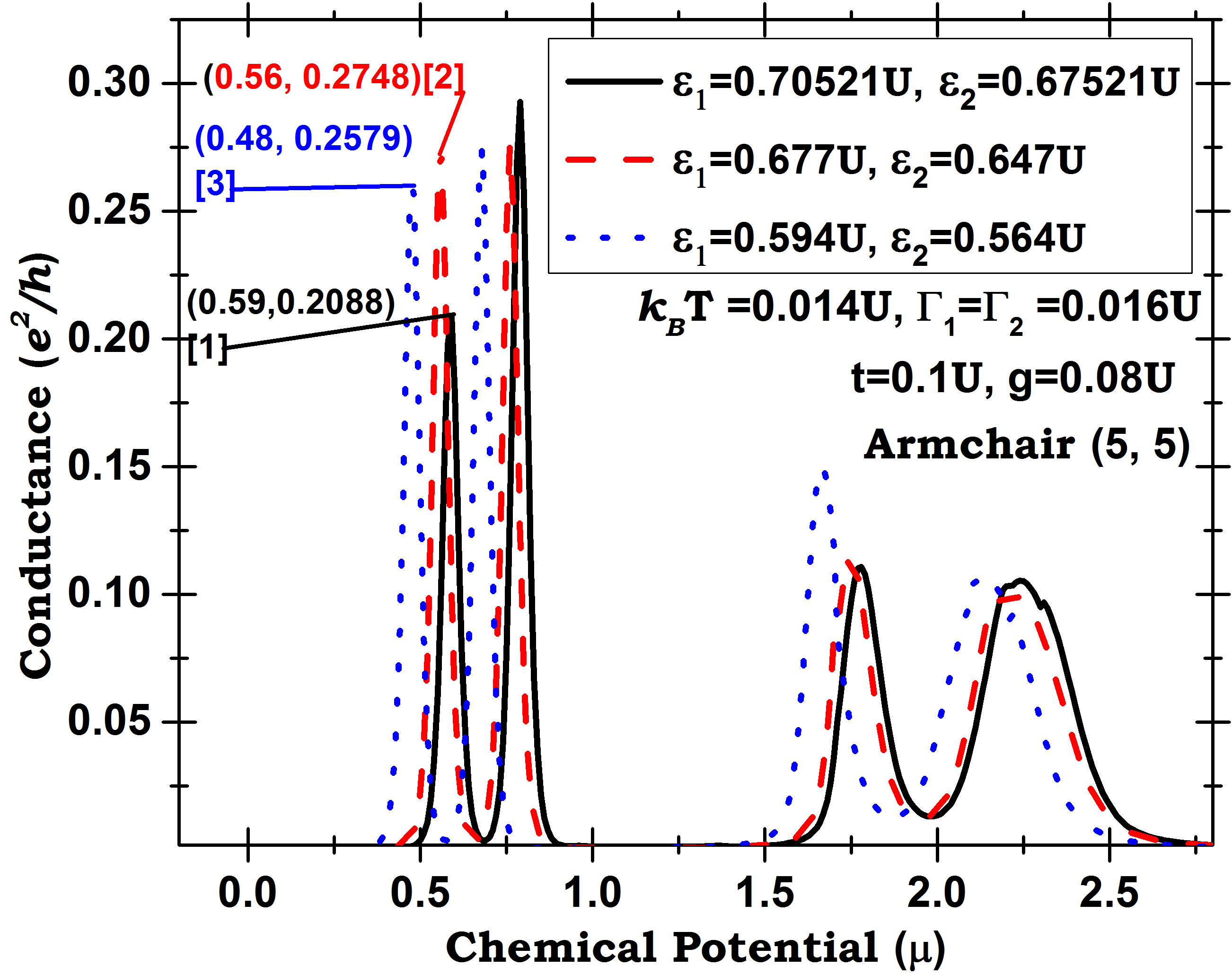} \label{singeffect1}}
\subfigure[\, $\varepsilon_{1}< \varepsilon_{2}$]{\includegraphics[width=0.9\linewidth]{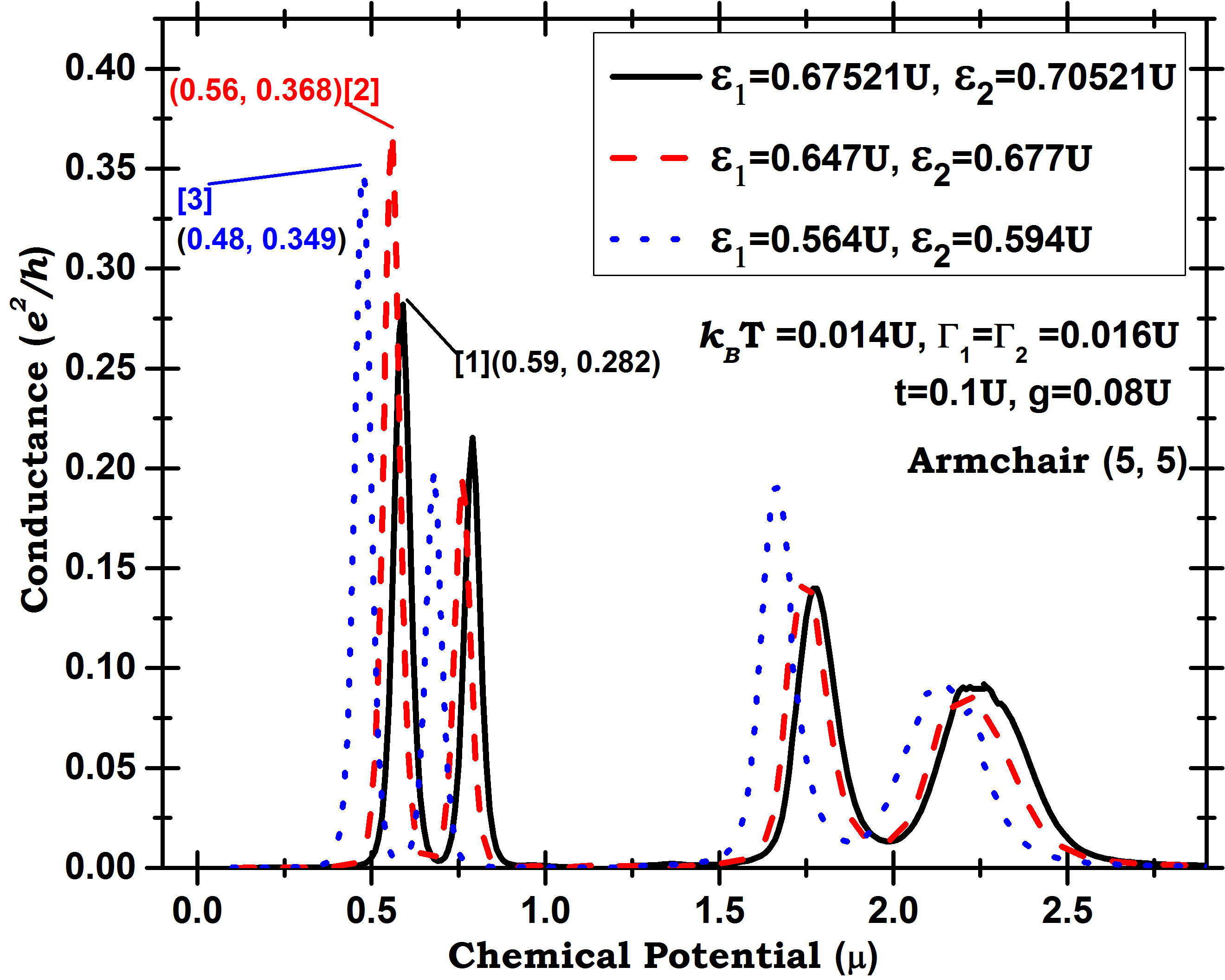} \label{singeffect2}}
\caption{Conductance (in units of $e^{2}/h$) {\it versus}  chemical potential $\mu$ (in units of $U$) for armchair $(5,5)$ CNT. The plots correspond to three different set of values of the energy levels of two dots, in (a) for $\varepsilon_{1}=0.70521U$, $\varepsilon_{2}=0.67521U$; $\varepsilon_{1}=0.677U$, $\varepsilon_{2}=0.647U$ and $\varepsilon_{1}=0.594U$, $\varepsilon_{2}=0.564U$. (b) for $\varepsilon_{1}=0.67521U$, $\varepsilon_{2}=0.70521U$; $\varepsilon_{1}=0.647U$, $\varepsilon_{2}=0.677U$ and $\varepsilon_{1}=0.564U$, $\varepsilon_{2}=0.594U$. Other parameters are taken as $t=0.1U$, $g=0.08U$, $k_{B}T=0.014U$ and $\Gamma_{1}=\Gamma_{2}=0.016U$.}
\label{singularityeffect}
\end{figure}
\\
\indent 
The study, so far has not revealed any distinctive attribute of the vHs. Now, we investigate the effect of one of the vHs present in the DOS of the armchair $(5,5)$ CNT leads. In Fig. {\ref{singularityeffect}}, we plot, at a fixed value of the interdot tunneling matrix-element $t=0.1U$ with $g=0.08U$, the conductance as a function of  chemical potential $\mu$ in the armchair $(5,5)$ CNT leads for three different set of values of the energy levels of two dots, in Fig. {\ref{singeffect1}}, for $\varepsilon_{1}=0.70521U$, $\varepsilon_{2}=0.67521U$; $\varepsilon_{1}=0.677U$, $\varepsilon_{2}=0.647U$ and $\varepsilon_{1}=0.594U$, $\varepsilon_{2}=0.564U$. Where in Fig. {\ref{singeffect2}}, for $\varepsilon_{1}=0.67521U$, $\varepsilon_{2}=0.70521U$; $\varepsilon_{1}=0.647U$, $\varepsilon_{2}=0.677U$ and $\varepsilon_{1}=0.564U$, $\varepsilon_{2}=0.594U$. The couplings, are taken even smaller $\Gamma_{1}=\Gamma_{2}=0.016U$, so that the many-body eigenstates of the DQD system are less affected. The temperature is kept as the smallest parameter $k_{B}T=0.014U <\Gamma_{1}(\Gamma_{2})$. 
\begin{table*}
\caption{Positions of the all four peaks and heights of the first peaks (denoted by [1],[2] and [3] correspond to solid, dashed and dotted lines, respectively) in the conductance profiles given in Figs. {\ref{singeffect1}} and {\ref{singeffect2}}.}
\begin{tabular}{c|c|c|c|c|c|c|c|c|c|c|c|c|}
\hline
\multicolumn{1}{|c|}{}                                                            & \multicolumn{4}{c|}{}                                                                                                                         & \multicolumn{8}{c|}{Actual positions of the conductance peaks}                                                                                                                      \\ \cline{6-13} 
\multicolumn{1}{|c|}{\multirow{-2}{*}{Line}}                                      & \multicolumn{4}{c|}{\multirow{-2}{*}{\begin{tabular}[c]{@{}c@{}}Positions of the peaks (using \\ eigenstates of the isolated DQDs)\end{tabular}}} & \multicolumn{4}{c|}{Fig. {\ref{singeffect1}}}                                                                        & \multicolumn{4}{c|}{Fig. {\ref{singeffect2}}}                                                \\ \hline
\multicolumn{1}{|l|}{}                                                            & \multicolumn{1}{l|}{$1^{st}$}      & \multicolumn{1}{l|}{$2^{nd}$}     & \multicolumn{1}{l|}{$3^{rd}$}     & \multicolumn{1}{l|}{$4^{th}$}    & $1^{st}$                     & $2^{nd}$ & $3^{rd}$                     & $4^{th}$                    & $1^{st}$                     & $2^{nd}$ & $3^{rd}$         & $4^{th}$        \\ \hline
\multicolumn{1}{|c|}{Solid}                                                       & \cellcolor[HTML]{C0C0C0}0.5891     & 0.8297                            & 1.7107                            & 1.9513                           & \cellcolor[HTML]{C0C0C0}0.59 & 0.79     & 1.78                         & 2.24                        & \cellcolor[HTML]{C0C0C0}0.59 & 0.79     & 1.77             & 2.22            \\ \hline
\multicolumn{1}{|c|}{Dashed}                                                      & 0.5609                             & 0.8015                            & 1.6825                            & 1.9231                           & 0.56                         & 0.76     & 1.75                         & 2.22                        & 0.56                         & 0.76     & 1.75             & 2.20            \\ \hline
\multicolumn{1}{|c|}{Dotted}                                                      & 0.4779                             & 0.7185                            & 1.5995                            & 1.8401                           & 0.48                         & 0.68     & 1.67                         & 2.14                        & 0.48                         & 0.68     & 1.66             & 2.14            \\ \hline
\multicolumn{1}{l|}{}                                                             & \multicolumn{6}{c|}{}                                                                                                                                                                   & \multicolumn{6}{c|}{}                                                                                                                     \\
\multicolumn{1}{l|}{\multirow{-2}{*}{}}                                           & \multicolumn{6}{c|}{\multirow{-2}{*}{Fig. {\ref{singeffect1}}}}                                                                                                                                     & \multicolumn{6}{c|}{\multirow{-2}{*}{Fig. {\ref{singeffect2}}}}                                                                                       \\ \hline
\multicolumn{1}{|c|}{\begin{tabular}[c]{@{}c@{}}Heights of the\\ first peak\end{tabular}} & \multicolumn{2}{c|}{\cellcolor[HTML]{C0C0C0}{[}1{]}=0.209}             & \multicolumn{2}{c|}{{[}2{]}=0.275}                                   & \multicolumn{2}{c|}{{[}3{]}=0.258}      & \multicolumn{2}{c|}{\cellcolor[HTML]{C0C0C0}{[}1{]}=0.282} & \multicolumn{2}{c|}{{[}2{]}=0.368}      & \multicolumn{2}{c|}{{[}3{]}=0.349} \\ \hline
\end{tabular}
\label{tab:peakspositions}
\end{table*} 
In order to find, how a vHs present in the SWCNT leads can affect the conductance profile, we have deliberately fixed the energy levels of the QDs in such a way that the one electron chemical potential of the  isolated DQD system is aligned with the vHs position in the leads. For this purpose, the set of values $(\varepsilon_1,\varepsilon_2)$ for energy levels of the QDs in Figs. {\ref{singeffect1}} and {\ref{singeffect2}}, are kept separated as $\left|\varepsilon_{1}-\varepsilon_{2}\right|=0.03U$, but the one electron ground state of the DQD system corresponds to only one such set for solid line curves. Table {\ref{tab:peakspositions}}, shows that only the positions of the first peaks corresponding to solid lines; are very close to the first vHs position present; away from the Fermi level in the DOS of the armchair $(5,5)$ CNT leads at $0.589$. As can be noted from Table {\ref{tab:peakspositions}}, that the heights of the first conductance peaks corresponding to solid lines are smaller as compare to those of dashed and dotted lines $\left(\footnotesize {\left[1\right]<\left[2\right]\left(\left[3\right]\right)}\right)$. This is the effect of vHs. As can be noted from Figs. {\ref{singeffect1}} and {\ref{singeffect2}}, the pair of values $(\varepsilon_1,\varepsilon_2)$ for dotted lines has one of the energy level of two dots fixed at $0.594U$, which is very close to the first vHs position, but the corresponding heights of the first peaks are not the shortest because in this situation one-electron chemical potential of the isolated DQDs does not align with the vHs position. Similar effects of the vHs have also been observed for armchair $(6,6)$. It is thus observed that, adjusting system parameters, the one-electron chemical potential of the DQD system can be made to align with the vHs in the CNT leads. In this situation, the probability amplitude of an electron (from leads) to occupy the DQD system decreases and gets delocalized over entire Source-DQD-Drain system. Since the DOS at the vHs positions are high the delocalization of probability amplitude lead to reduction in occupancies of the dots and hence fall in height of the conductance peak.
\begin{figure}[htb]
\centering
\includegraphics[width=0.45\textwidth]{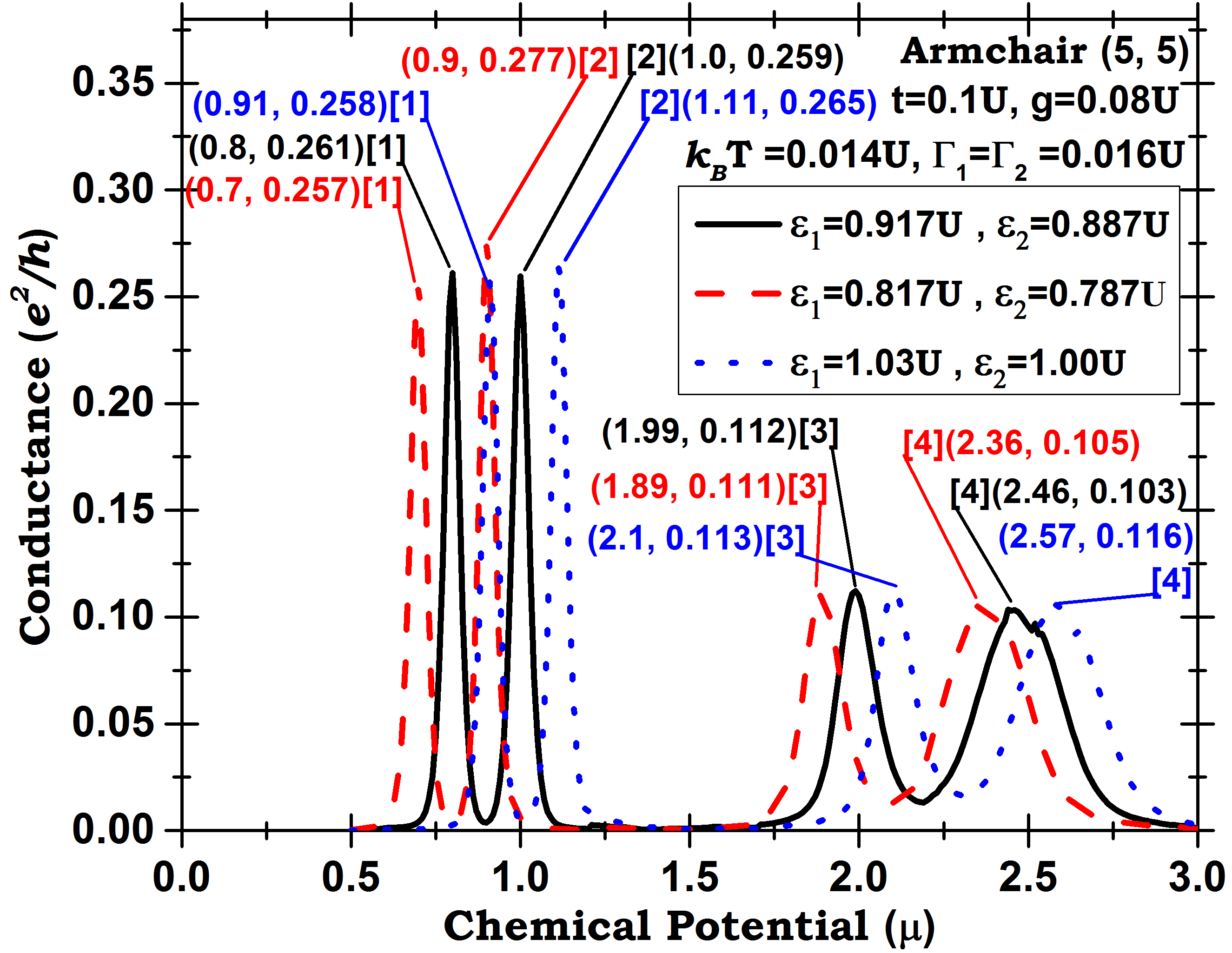}
\caption{{Conductance (in units of $e^{2}/h$) {\it versus}  chemical potential $\mu$ (in units of $U$) for armchair $(5,5)$ CNT. The plots correspond to three different set of values of the energy levels of two dots: $\varepsilon_{1}=0.917U$, $\varepsilon_{2}=0.887U-$solid line; $\varepsilon_{1}=0.817U$, $\varepsilon_{2}=0.787U-$dashed line and $\varepsilon_{1}=1.03U$, $\varepsilon_{2}=1.0U-$dotted line. Other parameters have been taken as , $g=0.08U$, $k_{B}T=0.014U$, $t=0.1U-$ and $\Gamma_{1}=\Gamma_{2}=0.016U$.\hfill \break}}
\label{SecondPeakFall}
\end{figure}
\\
\indent
For a situation when two-electron chemical potential aligns with one of the vHs position, we plot in Fig. {\ref{SecondPeakFall}, the conductance for three different set of values of the energy levels $\left(\varepsilon_{1}, \varepsilon_{2}\right)$ of QDs. 
The parameters are taken as $\varepsilon_{1}>\varepsilon_{2}$ such that the difference $\left|\varepsilon_{1} - \varepsilon_{2}\right|=0.03U$ and other parameters are the same as in Fig. {\ref{singularityeffect}}. The set of values $\varepsilon_{1}=0.917U$, $\varepsilon_{2}=0.887U$ corresponding to solid curve are adjusted such that the two electron chemical potential of the isolated DQD system coincides with vHs present in DOS of armchair $(5,5)$ CNT at 1.0. For solid curve the positions of all four conductance peaks obtained using the eigenvalues of the isolated DQDs are respectively given as $0.801$, $1.042$, $1.922$ and $2.163$. From the results given in Figs. \ref{singatdot1}, {\ref{sing2atdot1}, \ref{sing1e1ge2} and \ref{singeffect1}, we find that when $\varepsilon_{1}>\varepsilon_{2}$ the height of the second conductance peak is generally higher than that of the first peak. But, as can be noted that the height of the second conductance peak corresponding to the solid curve is less than that of the first peak whereas the heights of the second conductance peaks are higher than that of their corresponding first peaks for dashed and dotted curves. This observation is similar to the one observed in Fig. {\ref{singularityeffect}}.    
\begin{figure}
\subfigure[\, Conductance in the absence of vHs at $0.589$]{\includegraphics[width=0.9\linewidth]{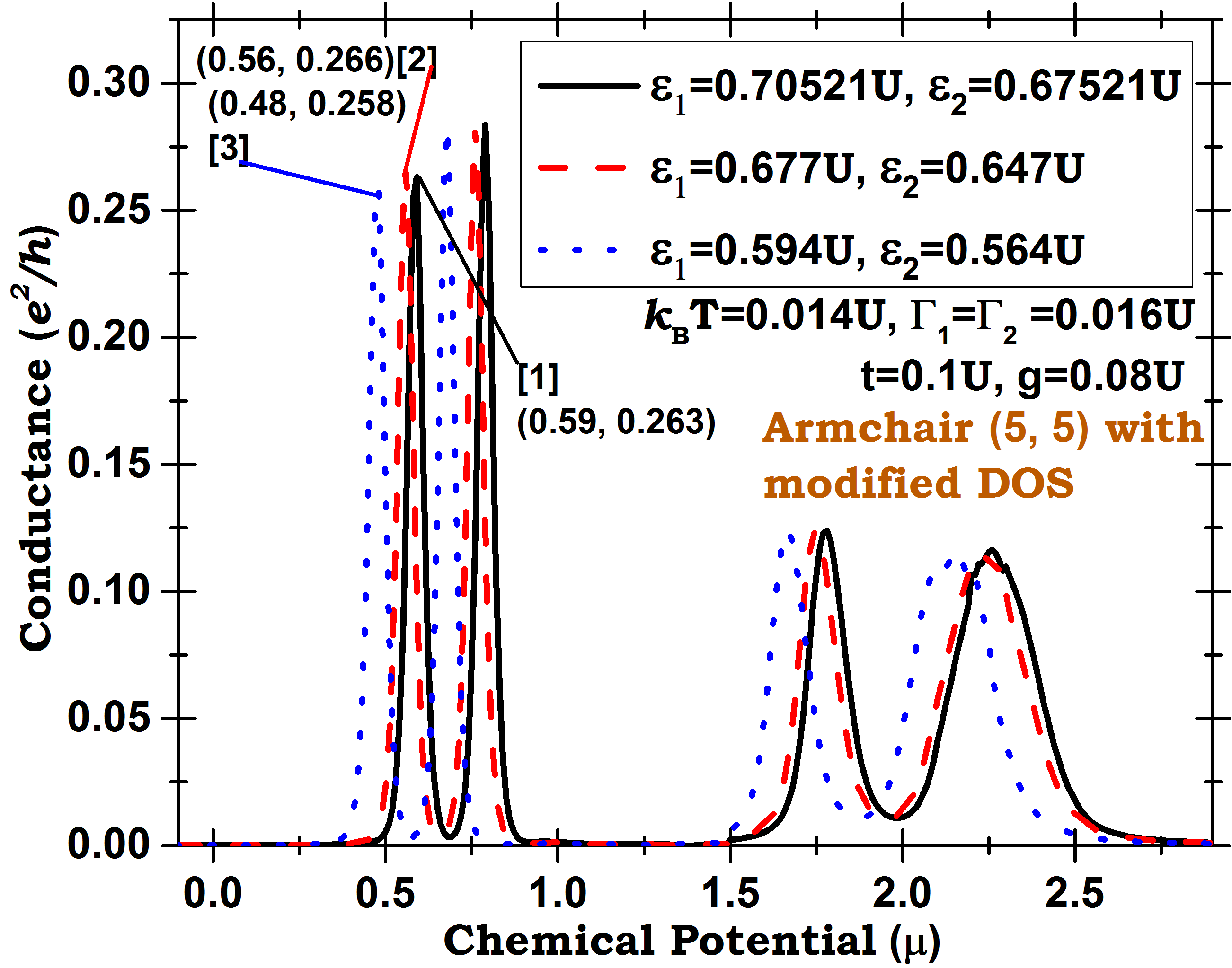} \label{resultWithmodifiedDos}}
\subfigure[\, DOS of armchair $(5,5)$ {\it versus} Modified DOS in the absence of vHs at $0.589$]{\includegraphics[width=0.9\linewidth]{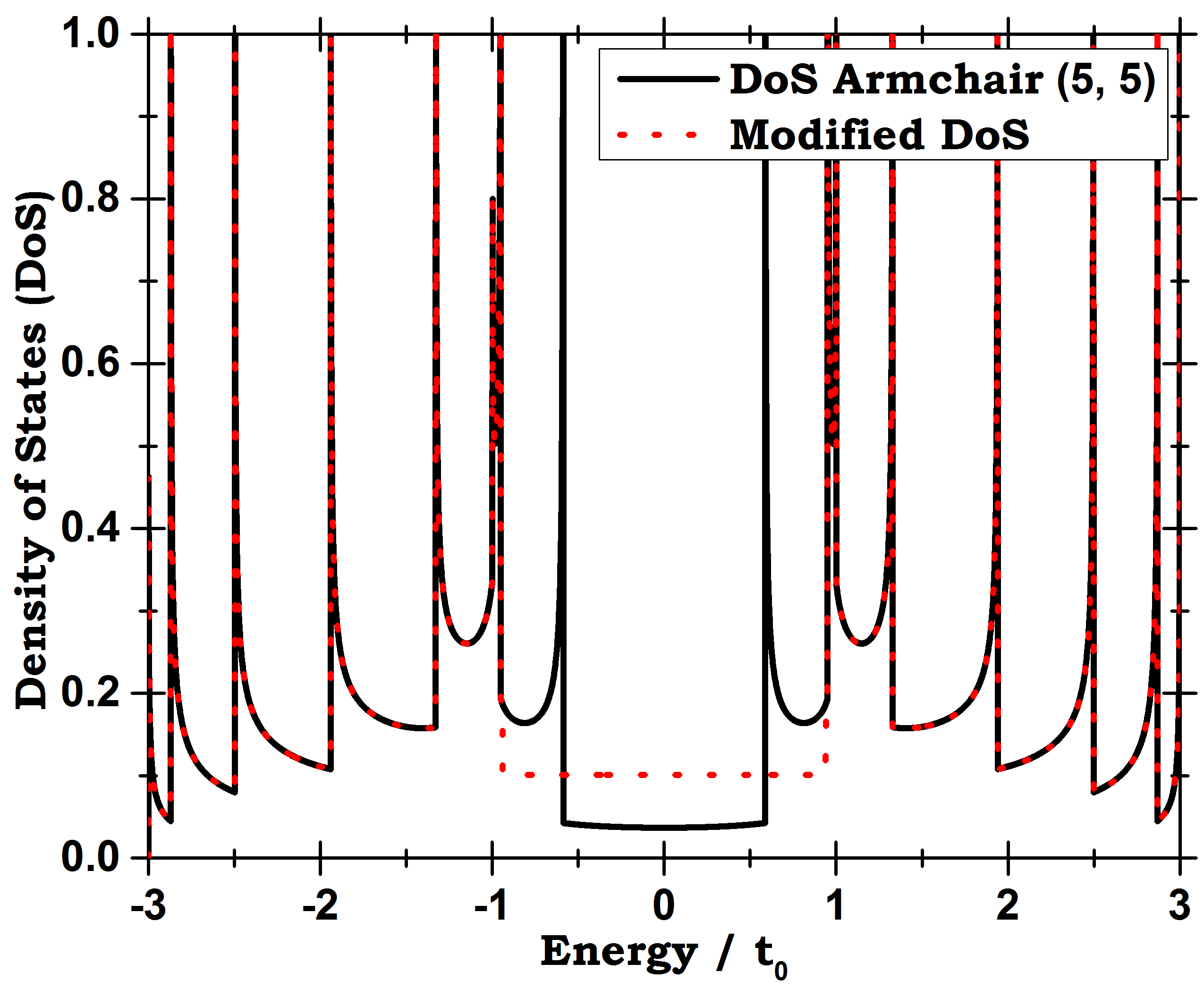} \label{OrigDoSvsModifiedDos}}
\caption{(a) Conductance (in units of $e^{2}/h$) {\it versus} chemical potential $\mu$ (in units of $U$) for modified DOS as given in Fig. {\ref{OrigDoSvsModifiedDos}}. The plots correspond to three different set of values of the energy levels of two dots: $\varepsilon_{1}=0.70521U$, $\varepsilon_{2}=0.67521U-$solid line; $\varepsilon_{1}=0.677U$, $\varepsilon_{2}=0.647U-$dashed line and $\varepsilon_{1}=0.594U$, $\varepsilon_{2}=0.564U-$dotted line. Other parameters are taken as $t=0.1U$, $g=0.08U$, $k_{B}T=0.014U$ and $\Gamma_{1}=\Gamma_{2}=0.016U$. (b) The DOS of armchair $(5,5)$ CNT and its modified form such that the first vHs at $E=0.5887$ is absent. The DOS is modified for $-0.94\leq E/t_{0}\leq 0.94$ in such a way that the original and modified DOS are normalized. \hfill \break}
\label{ModifiedDos}
\end{figure}
\\
\indent
In order to further investigate the observations in Figs. \ref{singularityeffect} and \ref{SecondPeakFall}, we plot in Fig. \ref{OrigDoSvsModifiedDos}, the modified DOS of the armchair $(5,5)$ CNT. To clarify whether the effect of vHs manifests itself as fall in the height of the conductance peak or this is some numerical artifact. The DOS is modified in such a way that the vHs at $0.589$ is absent but the positions of the other vHs remain unchanged and the area under the curve is also unchanged. In Fig. {\ref{resultWithmodifiedDos}}, we plot, the conductance as a function of the chemical potential $\mu$ for the same values of parameters as in Fig. {\ref{singeffect1}}. The conductance profiles comes out to be different because of the modified DOS of the leads. The height of the first peak for solid curve corresponding to the set of values $\varepsilon_{1}=0.70521U$, $\varepsilon_{2}=0.67521U$ does not fall shorter as there is no vHs present to coincide at $0.589$ as in Fig. \ref{singeffect1}. Therefore, the conductance peaks can significantly fall in heights when their positions are aligned with the vHs in the CNT leads.    
\section{Conclusion}
\label{coclude}
The DQD system with armchair $(5,5)$ SWCNT leads has been studied using Keldysh NEGF formalism. The effect of vHs in the DOS of CNT leads on the conductance has been examined. The conductance profiles with armchair $(5,5)$ CNT leads are qualitatively similar to those with the ideal leads. However, the heights of the conductance peaks with CNT leads are significantly higher. \\
\indent
The presence of vHs in DOS of CNT leads significantly affect the heights of the conductance peaks. It is observed for the case of one or two-electron chemical potential of the isolated DQD that when it align with the vHs in the DOS of CNT leads, the height of the corresponding conductance peak falls considerably. The effect of first vHs close to the Fermi level was found to prominently affect the height of the first conductance peak corresponding to non-interacting case. The effect of vHs is due the fact that whenever the chemical potential of the isolated DQD align with the vHs in the CNT leads the eigenstate of the system delocalized over entire source-DQD-drain system and the probability amplitude of an electron to occupy the DQD system decreases. In order to enlighten  the effect of vHs we have modified the DOS of armchair $(5,5)$ CNT such that the first vHs is absent and remaining vHs are present at the same positions. With same parameters as in the case when first vHs causes fall in the conductance peak height, it is found that the corresponding peak height is now unaffected.
\\
\indent
The heights of the conductance peaks can also be affected for other reasons. For example, when energy levels of the DQDs are interchanged for $\varepsilon_{1}\neq\varepsilon_{2}$, the conductance profiles also changes because the Hamiltonian of the system is not $1\leftrightarrow 2$ symmetric (where $1,\, 2$ labels the dots). 
It is observed that the heights of the alternate conductance peaks are suppressed even when any or both of the two energy levels of the dots align with the vHs position in the leads, the suppressions in the heights of the alternate peaks depends on following two factors. First, it depends on relative positions of the energy levels of the dots {\it i.e.} when $\epsilon_{1}>\varepsilon_{2}$, the first and third peaks are suppressed in their heights and when $\epsilon_{2}>\varepsilon_{1}$, the second and fourth peaks are suppressed. Second, the suppression effect increases with increasing the separation between the energy levels of the dots $\left|\epsilon_{1}-\varepsilon_{2}\right|$. 
\\  
\indent
The role of the interdot Coulomb interaction $g$ is significant when there are an average number of three or four electrons in the system this is reflected from the facts that the heights and widths of the third and fourth peaks are more affected as compared to the first and second peaks. In case of two electrons the effect due to interdot Coulomb interaction $g$ is very small as the two electron can occupy different dots.

\end{document}